\documentclass[lettersize,journal]{IEEEtran}
\usepackage{amsmath,amsfonts}
\usepackage{algorithmic}
\usepackage{algorithm}
\usepackage{array}
\usepackage{url}
\usepackage{textcomp}
\usepackage{stfloats}
\usepackage{url}
\usepackage{makecell}
\usepackage{multirow}
\usepackage{verbatim}

\usepackage{subfigure}
\usepackage{cite}
\usepackage{fancyhdr}
\usepackage{graphicx}
\usepackage{epstopdf}
\usepackage{soul}
\usepackage{xcolor}

\def\BibTeX{{\rm B\kern-.05em{\sc i\kern-.025em b}\kern-.08em
    T\kern-.1667em\lower.7ex\hbox{E}\kern-.125emX}}
\usepackage{fancyhdr}
\fancypagestyle{firststyle}{
	\fancyhf{}
	\fancyhead[L]{H. Poddar (Sharp), D. Gold (Nokia), D. Lee (Intel), N. Zhang (ZTE), G. Sridharan (Qualcomm), H. Asplund (Ericsson) and M. Shafi (Spark NZ), ``Overview of 3GPP Release 19 Study on Channel Modeling Enhancements to TR 38.901 for 6G'', \textit{Submitted in IEEE Vehicular Technology Magazine (VTM)}, July 2025, pp. 1--9.}

}
\begin{document}
\title{%Overview of 3GPP Release 19 Channel Modeling Study on 7-24 GHz, Near-Field Propagation, and Spatial Non-Stationarity for 6G
Overview of 3GPP Release 19 Study on Channel Modeling Enhancements to TR 38.901 for 6G
% \thanks{ This work is
% supported by xxxxx.}
}
\author{\IEEEauthorblockN{Hitesh Poddar, Dimitri Gold, Daewon Lee, Nan Zhang, Gokul Sridharan, Henrik Asplund, Mansoor Shafi}
}
\maketitle

% Comment this if you dont want the fancy header on top 
\thispagestyle{firststyle}

\begin{abstract}

Channel models are a fundamental component of wireless communication systems, providing critical insights into the physics of radio wave propagation. As wireless systems evolve every decade, the development of accurate and standardized channel models becomes increasingly important for the development, evaluation and performance assessment of emerging technologies. An effort to develop a standardized channel model began around 2000 through the Third Generation Partnership Project (3GPP) and the International Telecommunication Union (ITU) with the aim of addressing a broad range of frequencies from sub-1 GHz to 100 GHz. Prior efforts focused heavily on sub-6 GHz bands and mmWave bands, and there exist some gaps in accurately modeling the 7–24 GHz frequency range, a promising candidate band for 6G. To address these gaps, 3GPP approved a Release (Rel) 19 channel modeling study. This study resulted in several enhancements to the channel models, including the ability to accurately model a Suburban Macrocell (SMa) scenario, realistic User Terminal (UT) antenna models, variability in the number of clusters, variability in the number of rays per cluster, a framework for capturing variability in power among all polarizations, near field (NF) propagation, and spatial non-stationarity (SNS) effects, all of which may be crucial for future 6G deployments.  This paper presents the outcomes of this study and provides an overview of the underlying rationale, and key discussions that guided the validation, refinement, and enhancements of the 3GPP TR 38.901 channel models.

\end{abstract}


\section{Introduction}
\begin{figure}[h!]
    \centering
    \includegraphics[width=\linewidth]{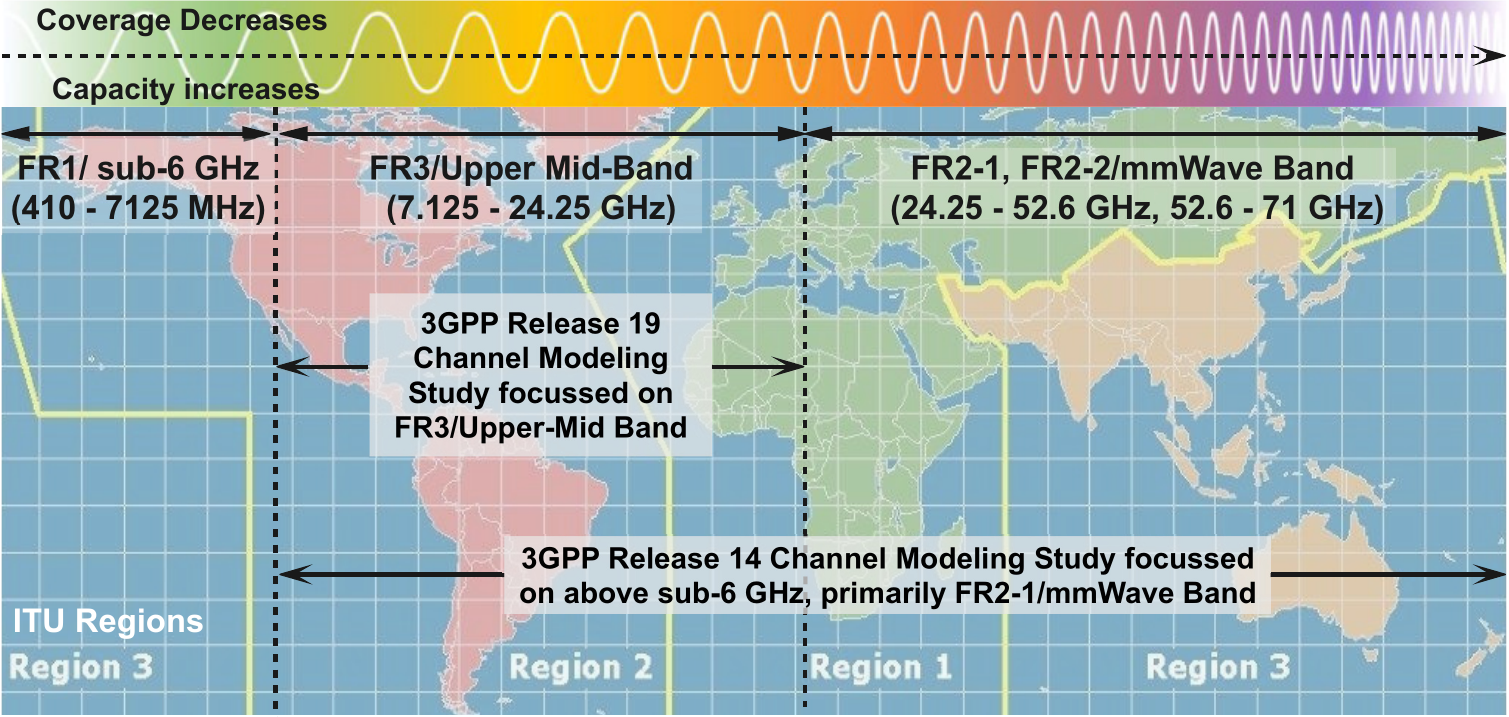}
    \caption{Overview of 3GPP channel modeling studies for IMT systems.}
    \label{fig:spectrum}
\end{figure}

%Channel models constitute a fundamental component of wireless communication system design. Prior to the deployment and operation of commercial radio systems, it is essential to develop a thorough understanding of radio wave propagation to create channel models that not only accurately capture key propagation mechanisms but also provide a framework for equitable comparison among different systems.
As the number of wireless devices continues to grow and their demand for higher data rates increases, it results in spectrum scarcity. Thus, additional spectrum is required for each new generation of cellular technology. Before costly spectrum auctions take place, it becomes essential to investigate and understand the radio propagation characteristics in these new frequency bands. Standardized channel models such as 3GPP TR 38.901 \cite{tr38901, tr38901v19}, ITU-R M.2412 \cite{2412} and ITU-R M.2135-1\cite{2135}, derived from extensive field measurements and simulations, have been instrumental in guiding the development of successive generations of wireless technologies. Channel modeling studies are initiated early in the standardization process, such as within 3GPP, to support the successful development of each generation of cellular technology in new spectrum ranges, as illustrated in Fig. \ref{fig:spectrum}. Furthermore, the ITU World Radiocommunication Conference (WRC) identifies key frequency bands for the global deployment of each new generation of International Mobile Telecommunications (IMT) systems. In addition to the existing sub-6 GHz and mmWave bands that will be reused for future 6G deployments, at ITU WRC 2023, several key frequency bands were identified as potential candidates for IMT 2030, or 6G which include \cite{ghosh:2024:world-radiocommunications}:
\begin{itemize}
\item 4.4-4.8 GHz, or parts thereof, in ITU Regions 1 and 3.
\item 7.125-8.4 GHz, or part thereof, in ITU Regions 2 and 3.
\item 7.125-7.25 GHz and 7.75-8.4 GHz, or part thereof, in ITU Region 3.
\item 14.8-15.35 GHz.
\end{itemize}
% All the candidate bands identified in ITU WRC 2023 are currently heavily utilized by a variety of incumbent systems. Thus, co-existence studies between future 6G systems and existing incumbents are critical for the roll-out and success of 6G. 
%currently being conducted in the ITU-R. %These studies will also significantly benefit from advancements in improved channel models.
 %Studying radio wave propagation and developing accurate channel models to capture its characteristics are inherently time-consuming, costly, and non-trivial tasks. Channel models are typically based on deterministic and stochastic parameters, derived from extensive measurements and simulations conducted over long periods across diverse scenarios. 
 \par As noted by the late Larry Greenstein (1937–2018), % a prominent figure in the field of radio propagation and 
 who is widely regarded as the father of radio propagation research \cite{goldsmith2018memory}
\textit{``Every time a new system has been built in a new band, in a new environment, or for a new service major questions have had to be answered about the nature of the radio propagation. It was true for Marconi’s wireless telegraph; it is true for today’s cellular systems; and it will be true for as long as people dream up new ways to use radio waves. Propagation is different at 6 GHz than at 850 MHz; indoor propagation differs from outdoor propagation; fixed wireless paths differ from mobile ones; and so on''}. This is again the case with the anticipated introduction of new spectrum for 6G, as it was in previous generations of cellular technologies and will continue to be in the future. To ensure the availability of accurate channel models, a 3GPP Rel-19 study item (SI) on channel modeling enhancements for 7-24 GHz for NR \cite{234018} was initiated in December 2023 and concluded in June 2025. %This has resulted in updates to \cite{tr38901}, %which supports channel modeling for frequencies ranging from 0.5 GHz to 100 GHz, as illustrated in Fig. \ref{fig:spectrum}. 
%We have also noted other modeling efforts in mid band. The IEEE 1944 Working Group completed the first meeting of the Extended mid-band frequencies Subgroup in November 2024, aiming to identify the gaps in sub-24 GHz frequency models, particularly in model parameters. This initiative aims to present an accurate model of the mid-band channels.
\par In this paper, we present a summary of the key enhancements, along with an overview of the underlying rationale, and key technical discussions that guided the validation, refinement, and enhancements of \cite{tr38901}, that are incorporated into \cite{tr38901v19}. Section \ref{sec:si_901} presents the scope of the 3GPP Rel-19 SI, followed by Section \ref{sec:major_discussion}, which discusses the various channel modeling components studied and incorporated \cite{tr38901v19}. These include the addition of a new SMa scenario (Subsection \ref{sec:sma}), realistic UT antenna models (Subsection \ref{sec:ue_ant_model}), refinements and new additions to channel model parameters (Subsection \ref{sec:chan_mod_params}), a framework for modeling variability in power among different polarizations (Subsection \ref{sec:polarization}), NF propagation (Subsection \ref{sec:nf}), and SNS effect (Subsection \ref{sec:sns}). Finally, conclusions are presented in Section \ref{sec:conclusion}.
%As of the writing of this paper, the standardization process remains ongoing. Significant structural changes to the model described in \cite{tr38901} are anticipated. For example, a new power allocation approach for rays within a cluster is needed to predict sparsity for the present model. Furthermore, the cross-polarization ratio (XPR) is an important performance metric that describes the relative power of a signal in orthogonal polarization directions compared to their co-polarized counterparts. A high absolute value of XPR indicates that the signal power in the orthogonal polarization direction is significantly lower than that in the co-polarized direction, which can help mitigate interference and enhance system performance. The XPR is significantly frequency selective in the frequency domain. Therefore, it is important to discuss XPR in the mid-band frequency range.
%However, this assumption lacks validation through empirical measurements. Additionally, a new suburban macrocell environment may need to be defined. This paper provides an overview of the current channel modeling studies in 3GPP and, due to space limits, presents only selected example measurements.

\section{3GPP Rel-19 SI on Channel Modeling}\label{sec:si_901}
%%%%%%%%% Description of the section %%%%%%%%%%%%%%%%%
% overview of TR 38.901, scope and motivation of study item
% Lead: Daewon
%%%%%%%%%%%%%%%%%%%%%%%%%%%%%%%%%%%%%%%%%%%%%%%%%%%%%%%
With 3GPP Rel-20 (6G studies) commencing in August 2025, the need of an accurate channel model has become essential. As the 3GPP TR 38.901 \cite{tr38901} channel models serve as industry-standard benchmarks for system design and evaluation, timely updates are critical for enabling 6G use cases, particularly in the 7-24 GHz frequency range. \cite{tr38901}, provides a comprehensive channel model for frequencies between 0.5-100 GHz for various scenarios such as Urban Microcell (UMi), Urban Macrocell (UMa), Rural Macrocell (RMa), Indoor Office (InH) and Indoor Factory (InF). However, more than 80\% of the data used for developing \cite{tr38901} was based on sub-6 GHz and above-24 GHz frequency bands, leaving the 6-24 GHz frequency range significantly underrepresented. Furthermore, \cite{tr38901} lacked the capability to accurately capture certain key propagation phenomena, such as NF propagation and SNS effects due to the deployment of extremely large antenna arrays (ELAAs). 
\par To address these limitations, in December 2023, 3GPP approved a Rel-19 SI \cite{234018}. The primary objectives of this SI are to validate the stochastic channel model in \cite{tr38901} for the 7-24 GHz frequency range using real-world measurement data for UMi, UMa, RMa, InH and InF scenarios and to adapt and extend the channel model in \cite{tr38901} to capture NF propagation, SNS effects and to develop frameworks for additional channel modeling components, scenarios, and other relevant aspects for next-generation wireless networks. This SI \cite{234018} was successfully concluded in June 2025.

The key outcomes of the SI that were incorporated into the updated TR 38.901 \cite{tr38901v19} are as follows:
\begin{itemize}
    \item Introduction of a new SMa scenario.
    \item Addition of a new realistic antenna model for handheld UTs and Consumer Premise Equipment (CPE).
    \item Inclusion of Absolute Time of Arrival modeling for UMi, UMa, InH and RMa scenarios.
    \item Implementation of a mechanism to modify the number of intra-cluster rays for scenarios involving large bandwidths and/or ELAAs. 
    \item Introduction of modeling variable number of clusters per Base Station (BS) and UT link.
    \item Framework for modeling variability in power among different polarizations.
    \item Development of NF propagation modeling to capture the characteristics of the spherical wavefront.
    \item Integration of SNS effect modeling to account for antenna element-wise power variation at both the BS and UT.
\end{itemize}
\renewcommand{\arraystretch}{1.15}
\begin{table*}[htbp]
  \centering
  \caption{Measurements (M) and Simulation (S) results provided in 3GPP Rel-19  SI for different environments (Env.), Frequencies (Freq.) and Bandwidths (BW) by various sources for validating, refining and enhancing the 3GPP TR 38.901 channel models. ``NA" indicates that the values were not reported. ``CW" indicates continuous wave.}
    \begin{tabular}{|p{1.6em}|p{8em}|p{1.5em}|p{6.5em}|p{6em}|p{30.415em}|}
    \hline
    \multicolumn{1}{|p{1.6em}|}{\textbf{Env.}} &
      \textbf{Source} &
      \textbf{Type} &
      \textbf{Freq. (GHz)} &
      \textbf{BW (GHz)} &
      \textbf{Results}
      \\
    \hline
    \multicolumn{1}{|c|}{\multirow{14}[10]{*}{UMi}} &
      Sharp, Nokia, NYU &
      M &
      6.75, 16.95  &
      1  &
      PL, DS, ASA, ASD, ZSA, ZSD
      \\
\cline{2-6}     &
      Nokia, Anritsu &
      M &
      3.5, 11, 29  &
      2  &
      Excess PL, DS
      \\
\cline{2-6}     &
      Intel &
      M &
      10  &
      0.25 &
      PL, DS
      \\
\cline{2-6}     &
      China Telecom &
      M &
      7, 10, 13  &
      NA &
      PL, DS, ASA, ZSA
      \\
\cline{2-6}     &
      Samsung &
      M &
      6.5, 13.5  &
      0.1 &
      PL, DS, ASA, ASD
      \\
\cline{2-6}     &
      AT\&T &
      M &
      7, 8, 15  &
      0.4 &
      PL, DS, ASA,  ZSA
      \\
\cline{2-6}     &
      Apple &
      S &
      8  &
      NA &
      PL, DS, ASD, ASA, ZSA, number of clusters
      \\
\cline{3-6}     &
      \multicolumn{1}{l|}{} &
      M &
      13  &
      0.55 &
      DS, ASA, ASD, ZSA, ZSD
      \\
\cline{2-6}     &
      ZTE &
      S &
      7, 8.4  &
      NA &
      NF, SNS
      \\
\cline{2-6}     &
      BUPT, Spark &
      M &
      \multicolumn{1}{l|}{13 } &
      \multicolumn{1}{l|}{0.4} &
      DS, ASD, ASA, ZSA, ZSD, number of clusters, number of rays per cluster, cluster: DS, ASD, ASA, ZSA, shadowing
      \\
\cline{2-6}     &
      Keysight &
      M &
      10.1  &
      0.5 &
      PL, DS, Rician K-factor, polarization, ASA, ZSA, ASD, ZSD
      \\
\cline{2-6}     &
      Ericsson &
      M &
      0.8625, 2.011, 5.02, 10.297, 22.001, 37  &
      CW &
      Excess PL
      \\
\cline{2-6}     &
      Huawei, HiSilicon &
      S &
      6.7, 10  &
      NA &
      SNS, NF
      \\
\cline{3-6}     &
      \multicolumn{1}{l|}{} &
      M &
      10  &
      0.5 &
      DS, ASD, ASA, ZSD, number of clusters, absolute time of arrival
      \\
    \hline
    \multicolumn{1}{|c|}{\multirow{16}[5]{*}{UMa}} &
      Samsung &
      M &
      6.5, 10.5, 13.5  &
      0.1 &
      ASD, ASA
      \\
\cline{2-6}     &
      AT\&T &
      M &
      7, 8, 15  &
      0.4 &
      PL, DS, ASA,  ZSA
      \\
\cline{2-6}     &
      Apple &
      M &
      13  &
      0.1 &
      PL, DS
      \\
\cline{3-6}     &
      \multicolumn{1}{l|}{} &
      S &
      8  &
      NA &
      PL, DS, ASD, ASA, ZSA, number of clusters
      \\
\cline{2-6}     &
      BUPT, Spark &
      M &
      \multicolumn{1}{l|}{13 } &
      \multicolumn{1}{l|}{0.4} &
      DS, ASD, ASA, ZSA, ZSD, number of clusters, number of rays per cluster, cluster: DS, ASD, ASA, ZSA, shadowing; NF,  SNS
      \\
\cline{2-6}     &
      BT, Ericsson &
      M &
      3.56, 3.7  &
      0.04 &
      ASD, ZSD
      \\
\cline{2-6}     &
      Vodafone, Ericsson &
      M &
      3.4  &
      0.1 &
      ASD, ZSD
      \\
\cline{2-6}     &
      Ericsson &
      M &
      3.5  &
      0.1 &
      DS, ASD, ZSD, number of clusters, SNS
      \\
\cline{3-6}     &
      \multicolumn{1}{l|}{} &
      M &
      1.8775, 5.25  &
      CW, 0.2 &
      Polarization, fast fading characteristics
      \\
\cline{3-6}     &
      \multicolumn{1}{l|}{} &
      M &
      13, 28  &
      50, 0.0055 &
      ASD, ZSD, Excess PL
      \\
\cline{3-6}     &
      \multicolumn{1}{l|}{} &
      M &
      2.6, 13  &
      20, 0.0055 &
      Absolute time of arrival, polarization
      \\
\cline{3-6}     &
      \multicolumn{1}{l|}{} &
      M &
      0.8625, 2.011, 5.02, 10.297, 22.001, 37  &
      CW &
      Excess PL
      \\
\cline{3-6}     &
      \multicolumn{1}{l|}{} &
      S &
      7, 10, 15, 20, 24  &
      NA &
      NF
      \\
\cline{2-6}     &
      Huawei, HiSilicon &
      M &
      3.5, 6.5, 15  &
      NA &
      PL
      \\
\cline{3-6}     &
      \multicolumn{1}{l|}{} &
      M &
      6.5, 13, 15  &
      0.16, 0.4, 0.25 &
      PL, DS, ASD, ASA, ZSA, number of clusters
      \\
\cline{3-6}     &
      \multicolumn{1}{l|}{} &
      S &
      6, 7, 10  &
      NA &
      NF, SNS
      \\
    \hline
    \multicolumn{1}{|c|}{\multirow{8}[3]{*}{SMa}} &
      Nokia &
      M &
      28  &
      2$e^{-5}$ &
      O2I building penetration loss, PL
      \\
\cline{2-6}     &
      AT\&T &
      M &
      7, 8, 15  &
      0.4 &
      PL, DS, ASA, ZSA
      \\
\cline{2-6}     &
      Apple &
      S &
      8  &
      NA &
      PL, DS, ASD, ASA, ZSA, number of clusters
      \\
\cline{2-6}     &
      ZTE &
      S &
      7  &
      NA &
      PL, LOS probability
      \\
\cline{2-6}     &
      BT, Ericsson &
      M &
      3.56, 3.7  &
      0.04 &
      ASD, ZSD
      \\
\cline{2-6}     &
      Vodafone, Ericsson &
      M &
      3.4  &
      0.08 &
      ASD, ZSD
      \\
\cline{2-6}     &
      Ericsson &
      S &
      NA &
      NA &
      LOS probability with and without vegetation
      \\
\cline{3-6}     &
      \multicolumn{1}{l|}{} &
      M &
      0.8625, 2.011, 5.02, 10.297, 22.001, 37  &
      CW &
      Excess PL
      \\
    \hline
    \multicolumn{1}{|c|}{\multirow{2}[1]{*}{RMa}} &
      Qualcomm &
      M &
      3.4, 13  &
      CW &
      PL
      \\
\cline{2-6}     &
      BT, Ericsson &
      M &
      3.56, 3.7  &
      0.04 &
      ASD, ZSD
      \\
    \hline
    \multicolumn{1}{|c|}{\multirow{10}[4]{*}{InH}} &
      Sharp, Nokia, NYU &
      M &
      6.75, 16.95  &
      1  &
      PL, DS, ASA, ASD, ZSA, ZSD
      \\
\cline{2-6}     &
      Rohde \& Schwarz &
      M &
      14  &
      2  &
      PL, DS, AS, number of rays, Rician K-factor
      \\
\cline{2-6}     &
      Sony &
      M/S &
      15  &
      0.1 &
      PL, DS
      \\
\cline{2-6}     &
      AT\&T &
      M &
      7, 8, 11, 15  &
      0.4 &
      PL, DS, ASA,  ZSA
      \\
\cline{2-6}     &
      Apple &
      M &
      13  &
      0.55, 0.1 &
      DS, PL
      \\
\cline{2-6}     &
      ZTE &
      M &
      6-10  &
      NA &
      DS, NF
      \\
\cline{2-6}     &
      BUPT, Spark &
      M &
      \multicolumn{1}{l|}{13 } &
      \multicolumn{1}{l|}{0.4} &
      DS, ASD, ASA, ZSA, ZSD, number of clusters, number of rays per cluster, cluster: DS, ASD, ASA, ZSA; NF and SNS
      \\
\cline{2-6}     &
      Keysight &
      M &
      10.1  &
      0.5  &
      PL, DS, Rician K-factor, polarization, ASA, ZSA, ASD, ZSD
      \\
\cline{2-6}     &
      Ericsson &
      S &
      7, 10, 15, 20, 24  &
      NA &
      NF
      \\
\cline{2-6}     &
      Huawei, HiSilicon &
      M/S &
      10, 6.7  &
      0.5, NA &
      DS, ASD, ASA, ZSD, number of clusters, absolute time of arrival, SNS
      \\
    \hline
    \multicolumn{1}{|c|}{\multirow{3}[1]{*}{InF}} &
      Sharp, Nokia, NYU &
      M &
      6.75, 16.95  &
      1  &
      PL, DS, ASA, ASD, ZSA, ZSD
      \\
\cline{2-6}     &
      Nokia, Anritsu &
      M &
      3.5, 11, 29  &
      2  &
      Excess PL, DS 
      \\
\cline{2-6}     &
      Apple &
      M &
      13  &
      0.1 &
      PL
      \\
    \hline
    \end{tabular}%
  \label{tab:meas}%
\end{table*}%
\renewcommand{\arraystretch}{1}
\section{Major Discussion}\label{sec:major_discussion}
%%%%%%%%% Description of the section %%%%%%%%%%%%%%%%%
% SI focused on 7-24  major discussion points including things that were not adopted/introduced for 9.8.1 and 9.8.2. will add a table summarizing all measurements submitted for Rel-19 NF and SNS (Hitesh).
% Lead: Daewon, Nan, Hitesh
%%%%%%%%%%%%%%%%%%%%%%%%%%%%%%%%%%%%%%%%%%%%%%%%%%%%%%%
%In the SI \cite{234018}, 
Global academic and industrial organizations contributed inputs on various channel modeling parameters, as shown in Table \ref{tab:meas}. These contributions span a range of key channel modeling parameters, such as path loss (PL), delay spread (DS), Azimuth Angular Spread of Arrival (ASA), Azimuth Angular Spread of Departure (ASD), Elevation Angular Spread of Arrival (ZSA), Elevation Angular Spread of Departure (ZSD), NF propagation and SNS effects. A detailed summary of the updated and newly introduced channel model parameters is provided in Table \ref{tab:updates}. Additionally, the following channel modeling parameters were considered validated based on provided data and no changes were made to the following:
\begin{itemize}
    \item PL model for UMi, InH, RMa and InF scenarios.
    \item DS for InH scenario.
    % \item Shadow fading for indoor office and urban micro non-line-of-sight (NLOS) deployments,
    \item ZSD for UMa scenario. 
\end{itemize}
Due to limited data and inconsistent validation results, the need to alter some channel modeling parameters remained inconclusive. Therefore, no changes were made to the following:
\begin{itemize}
    \item PL model for UMa scenario.
    \item DS for InF scenario.
    %\item Shadow fading for urban micro line-of-sight (LOS) deployments,
    \item ZSD for UMi scenario.
    \item ASA, ASD, ZSA and ZSD for InH scenario.
\end{itemize}

% Based on newly provided measurement data and previously published sources, updates were made to the following channel modeling parameters.
% \begin{itemize}
%     \item Delay spread for urban micro and urban macro deployments,
%     \item Azimuth angular spread for urban micro and urban macro deployments,
%     \item Elevation angular spread of arrival for urban micro and urban macro deployments,
%     \item Cluster angular spread of urban macro deployments.
% \end{itemize}
%In addition to validating and updating the existing channel modeling parameters, 
Moreover, the 3GPP Rel-19 SI also introduced several new channel modeling components, as detailed in the subsequent sections of this paper.

\subsection{SMa Scenario}\label{sec:sma}
%%%%%%%%% Description of the section %%%%%%%%%%%%%%%%%
% dicussions related to sma modelling - includes path loss, los, deployments, etc
% Lead: Gokul
%%%%%%%%%%%%%%%%%%%%%%%%%%%%%%%%%%%%%%%%%%%%%%%%%%%%%%%
Prior to the 3GPP Rel-19 SI, \cite{tr38901} included only three outdoor deployment scenarios: UMi, UMa and RMa. However, missing from this list was a scenario that more accurately reflects the characteristics of suburban deployments. These suburban scenarios are characterized by buildings that are typically low residential detached houses with one or two floors, or blocks of flats with a few floors. Occasional open areas such as parks or playgrounds between the houses make the environment rather open. Streets do not form urban-like regular strict grid structure and vegetation is more prevalent than in urban areas with a high variability of foliage density across different suburban areas. Additionally, the BSs are mounted well above rooftop heights, and generally higher line of sight (LOS) probabilities and lower PL compared to urban scenarios is observed. The Inter-site distances (ISDs) between the BSs in such areas typically range from 1200-1800 m. Thus, to address this gap, the SI introduced a SMa scenario and the key parameters used to characterize this scenario are summarized in Table \ref{SMa_basic_parameters}. The following aspects are considered:

\begin{itemize}
\item A new SMa LOS probability model is introduced based on ray tracing simulations. Given the significant variability in foliage, building density and heights across different suburban environments, the SMa LOS probability model adopted in Table 7.4.2-1 \cite{tr38901v19} follows an approach similar to the InF LOS probability model described in \cite{tr38901}. In contrast, the SMa LOS probability models used in \cite{winner2} and \cite{2135} do not account for such variations.
\item The SMa PL model in \cite{2135} is adopted from \cite{winner2} and validated using measurements submitted in 3GPP Rel-19 SI and reused without modifications in Table 7.4.1-2 \cite{tr38901v19}. However, their validity is extended up to 37 GHz based on new data, compared to 6 GHz in \cite{2135}. 
\item A new material penetration loss model for plywood is introduced in Table 7.4.3-1 \cite{tr38901v19}, based on measurements, as plywood is the most common construction material found in suburban residential homes.
\item In addition to the existing outdoor to indoor (O2I) building penetration loss model in \cite{tr38901}, a new O2I low loss model based on measurements conducted in the SMa scenario is introduced in Table 7.4.3-2 \cite{tr38901v19}. This model uses a composite of plywood and standard glass. In contrast, the existing O2I low loss models used a combination of concrete and glass, which is less representative of suburban residential buildings.
\item All fast fading model parameters are adopted from \cite{2135}, \cite{winner2}, or the UMa scenario in \cite{tr38901}. However, certain parameters such as DS, ASA, ZSA, ASD, and cluster ASD are derived using the arithmetic mean of corresponding values reported in \cite{2135} or \cite{winner2} and data provided in 3GPP Rel-19 SI.
\end{itemize}

\renewcommand{\arraystretch}{1.1}
\begin{table}[]
 \caption{Channel Model Parameters for SMa scenario}
     \label{SMa_basic_parameters}
    \centering
\begin{tabular}{|p{0.3\columnwidth} | p{0.56\columnwidth}|}
 \hline
\textbf{Parameter} & \textbf{Value} \\
\hline
 \vspace{2em} Cell layout & $\bullet$ Hexagonal grid, 19 macro sites, 3 sectors per site. \newline $\bullet$ Up to 2 floors for residential buildings, up to 5 floors for commercial buildings. \newline $\bullet$ Building distribution are 90\% residential and 10\% commercial buildings. \\
 \hline
 
ISD & 1200-1800 m  \\
 \hline
  BS antenna height & 35 m  \\
 \hline
 BS antenna downtilt & Mechanical downtilt of ${92^\circ-95^\circ}$ \\ 
\hline
 \vspace{0.8em} UT height & $\bullet$ 1.5 m for outdoor. \newline $\bullet$ 1.5 or 4.5 m for residential buildings. \newline $\bullet$ 1.5/4.5/7.5/10.5/13.5 m for commercial buildings. \\
\hline
UT ratio & 80\% indoor and 20\% outdoor \\
\hline
UT mobility & Indoor UTs: 3 km/h, outdoor UTs (in car): 40 km/h \\
\hline
Minimum BS-UT distance (2D) & 35 m \\
\hline
UT distribution (horizontal) & Uniform \\
\hline
UT distribution (vertical) & Uniform distribution across all floors for a building type \\
\hline
LOS probability & Table 7.4.2-1 in \cite{tr38901v19}\\
\hline
PL model & Table 7.4.1-2 in \cite{tr38901v19}\\
\hline
Plywood penetration loss model & Table 7.4.3-1 in \cite{tr38901v19}\\
\hline
O2I penetration loss model & $\bullet$ Table 7.4.3-2 in \cite{tr38901v19}
\newline $\bullet$ In-car penetration loss is applied to all outdoor UTs, Table 7.4.3.2 in \cite{tr38901v19}.\\
\hline
Fast fading model parameters & Table 7.5-6 Part 4 in \cite{tr38901v19}\\
\hline
Absolute Time of Arrival & Table 7.6.9-1 in \cite{tr38901v19}\\
\hline
    \end{tabular}
\end{table}
\renewcommand{\arraystretch}{0}

% The new outdoor to indoor (O2I) building penetration loss model that is applicable to indoor UTs uses a composite of plywood and standard glass. Plywood is a new material type introduced in Release 19 that differs slightly from the previously modeled wood. Note that the existing models used a combination of concrete and glass, which is less representative of suburban residential construction.

% Most of the large-scale parameters (LSPs) such as angular spreads and delay spreads for the SMa scenario are largely inherited from ITU-R M.2135-1 or are derived from new measurements contributed to Release 19 studies. For O2I users, due to the lack of sufficient measured data, most parameters are assumed to be similar to those from the non-line-of-sight (NLOS) conditions.

% The number of clusters for channel modeling is consistent with ITU-R M.2135-1: 15 clusters are used for LOS users, while 14 clusters are used for NLOS and indoor users.
\subsection{UT Antenna Modeling}\label{sec:ue_ant_model}
%%%%%%%%% Description of the section %%%%%%%%%%%%%%%%%
% dicussions related to the newly introduced UE antenna modeling in 901
% Lead: Dimitri
%%%%%%%%%%%%%%%%%%%%%%%%%%%%%%%%%%%%%%%%%%%%%%%%%%%%%%%
A model for the BS antenna panels is provided in Section 7.3 \cite{tr38901}. %These comprise directive single- or dual-polarized elements arranged in one or more sub-panels. 
In contrast, on the UT side, antenna modeling has historically remained oversimplified. \cite{tr38901} considers only isotropic UT antenna patterns with single or dual cross-polar components. For advanced scenarios, such as multiple-input multiple-output (MIMO) \cite{tr38.802} and FR2 studies \cite{tr38.803}, within 3GPP, more detailed UT antenna configurations were introduced. % For instance, in the Rel-14 study on NR , up to four UT antennas in FR1 and antenna arrays separated by half a wavelength were considered. Additionally,  also introduced UT antenna element directional radiation pattern for beamforming modeling. 
However, these assumptions remain unrealistic for handheld UTs due to:
\begin{itemize}
    \item Half-wavelength spacing of UT antennas leads to a frequency-independent combined radiation pattern.
    \item Real UT antennas exhibit certain directivity.
    \item The polarization components of the UT radiation pattern vary spatially around the device that is difficult to capture analytically. %\cite{R1-2501348}.
    \item There is a significant power imbalance between UT antennas, due to differences in their implementation and reception conditions.
\end{itemize}

\begin{figure}[t]
    \centering
    \includegraphics[width=\linewidth]{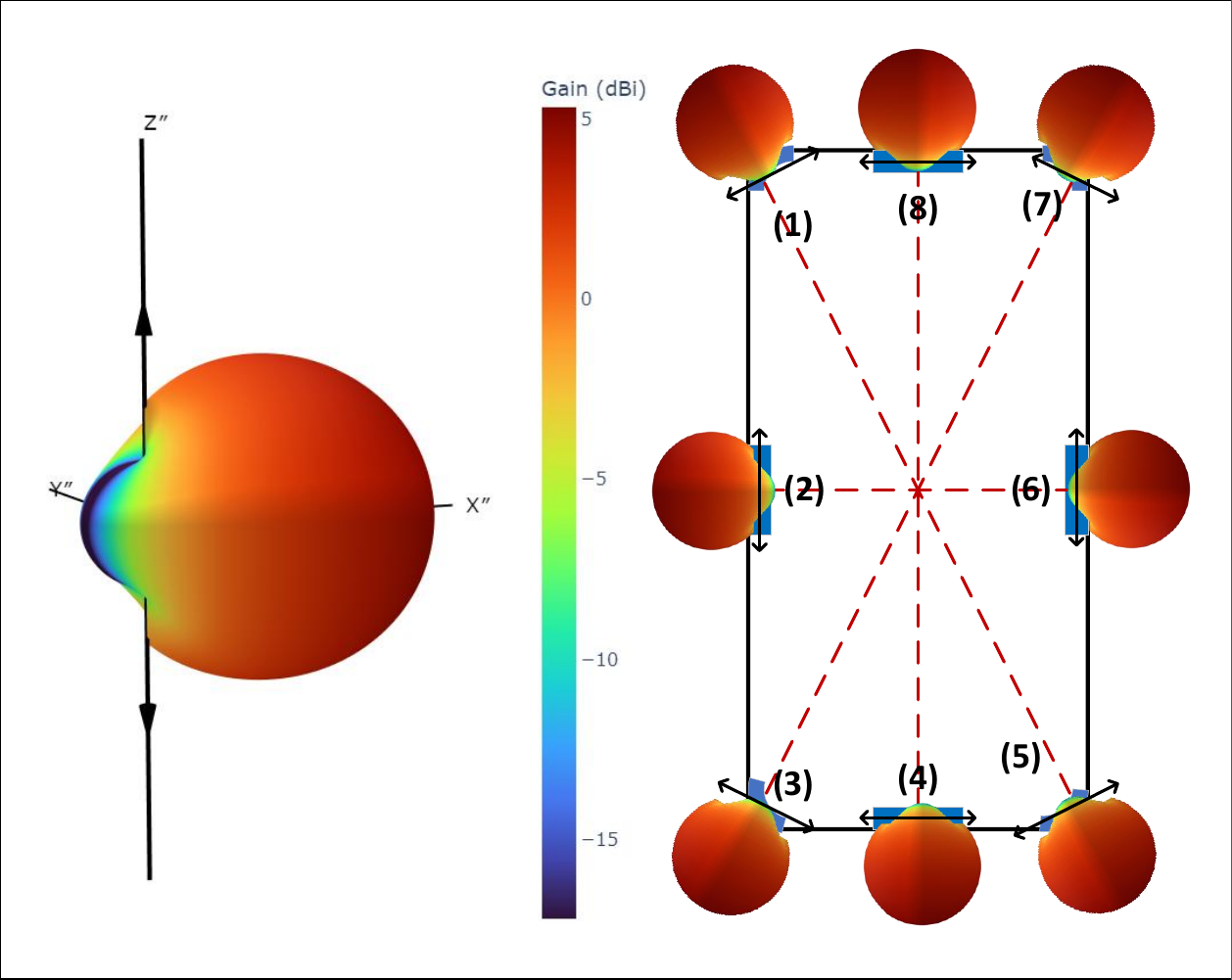}
    \caption{ UT antenna reference radiation pattern with 3 dB beamwidth of 125$^{\circ}$ and 5.3 dBi directional gain (left) and candidate antenna locations (right).}
    \label{fig:ue_antenna}
\end{figure}
To address these limitations, a more realistic UT antenna modeling framework was introduced in 3GPP Rel-19 SI. To obtain the frequency dependent combined UT radiation pattern, the physical dimensions of the UT was introduced. A reference UT size of 15 cm × 7 cm for handheld devices and 20 cm × 20 cm for CPE which imposes a fixed distance between antennas was adopted. Furthermore, eight candidate antenna locations for handheld UTs and nine for CPEs were identified, as shown in Fig. \ref{fig:ue_antenna} and Fig 7.3-6 \cite{tr38901v19}, respectively. %For calibration at around 7–8 GHz, it is assumed that only four antennas at candidate locations 1, 3, 5, and 7 are used. This assumption is reasonable due to the presence of other antennas in the UT, for example, those for frequencies below 1 GHz, FR2 antenna arrays, and antennas supporting non-3GPP technologies. At higher frequencies, e.g., 15 GHz and above, more candidate antenna locations can be simulated, including those for dual-polarized antennas or antenna arrays with multiple elements.
Each UT antenna is oriented along the vector from the device center to its location and the direction of maximum gain of the antenna are aligned with these directions. The new antenna pattern for the UT %$A''(\theta'',\phi'')$ 
is defined in Table 7.3-2 \cite{tr38901v19} and shown in Fig. \ref{fig:ue_antenna} on the left. % This allows direct performance comparisons with omnidirectional antennas. 
However, specifying only the maximum gain direction is insufficient to fully define the antenna radiation pattern orientation. Polarization must also be considered. When a single antenna field pattern is considered per location, the UT antenna reference radiated field is vertically polarized with all gain concentrated in the theta component. %: $F''_{\theta''}=\sqrt{A''(\theta'',\phi'')}$ and $F''_{\phi''}=0$.
This configuration is referred to as the polarization direction along the vertical $Z''$ axis. %in the reference antenna coordinate system.
When the reference radiation pattern is translated to the particular location, the maximum gain and the polarization directions are aligned as shown in Fig. \ref{fig:ue_antenna} on the right, where dashed lines indicate gain direction and arrows indicate polarization. Finally, the polarization components are transformed into the global coordinate system based on the orientation of the UT itself, using the coordinate transformation procedures defined in Section 7.1 \cite{tr38901v19}. As a result, both polarized receive field patterns $F_{rx,u,\theta}$ and $F_{rx,u,\phi}$ of UT antenna $u$ are generally non-zero, resulting in elliptical polarization. These components are used to compute the channel coefficients of the fast fading model described in section 7.5 \cite{tr38901v19}. Similar considerations apply when a candidate location supports two orthogonal polarization field patterns.
%In such cases, the patterns are orthogonal and additionally rotated by 45 degrees around the axis pointing from the UT center to the antenna.

To further improve the modeling accuracy, two new optional components allow simulation of power imbalance across UT antennas. The first model introduces a randomized loss per antenna port. This does not model the entire radio frequency chain, but accounts for variation in antenna performance due to placement, shape, and implementation differences. Although by default no imbalance is modeled, during 3GPP discussions, random sampling values per antenna ranging from -2 dB to + 3 dB were proposed\cite{R1-2503130}.
\begin{figure}[t]
    \centering
    \includegraphics[width=\linewidth]{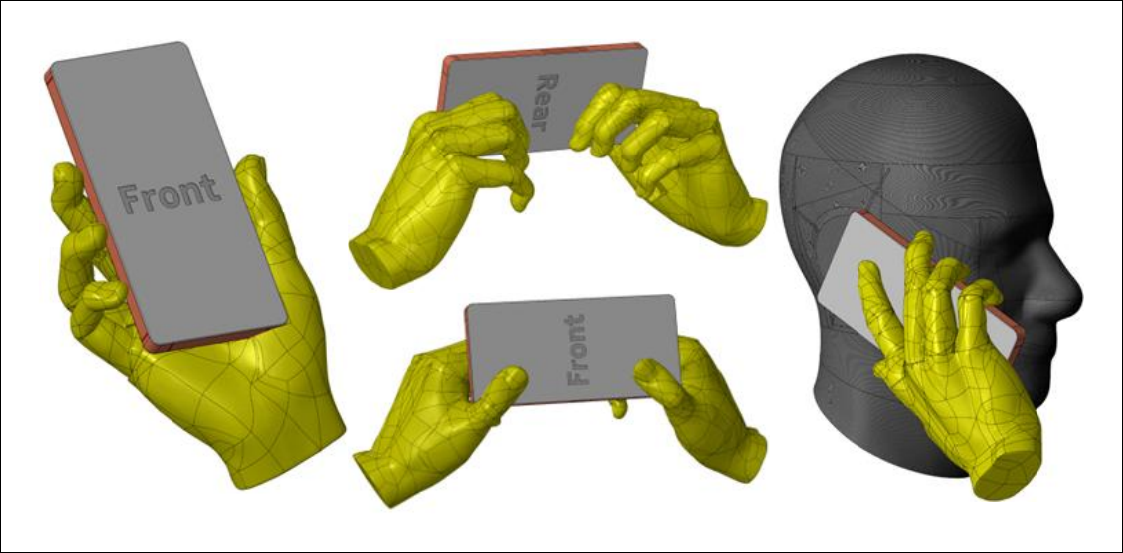}
    \caption{UT antenna blockage scenarios (left to right): one hand grip, dual hand grip, head and hand grip.}
    \label{fig:ue_grip}
\end{figure}
The second enhancement introduces a model to simulate SNS effect at the UT, accounting for power variations due to blockages from the user’s hand or head. \cite{tr38901} already defines two blockage models: Model A, which uses a stochastic approach including self-blocking, and Model B, which adopts a geometric method to capture human and vehicular blockage. Both models address far-field blockage effects, applying attenuation equally across all UT antennas. However, Cellular Telecommunications and Internet Association (CTIA)-based phantom model \cite{ctia2015phantom} simulations show that uniform attenuation across blocked angles deviates from real single-hand grip behavior. %\cite{R1-2501349}.
New UT SNS model assumes that 90\% of UTs are affected by blockage, with the remaining 10\% operating in free space. Three grip scenarios are modeled: single-hand grip, dual-hand grip, and head-and-hand grip as shown in Fig. \ref{fig:ue_grip}. Their occurrence probabilities are defined in Table 7.6.14.2-1 \cite{tr38901v19}. Each grip scenario affects the UT antennas differently. Attenuation values per antenna location are specified in Table 7.6.14.2-2 \cite{tr38901v19} based on simulations for frequencies up to 8.4 GHz %\cite{R1-2502900}, 
and based on measurements for frequencies around 15 GHz. %\cite{R1-2502327}. 
Incorporating antenna element-wise power variations due to SNS effect enables a more accurate performance evaluation of massive MIMO systems.
\subsection{Channel Model Parameters} \label{sec:chan_mod_params}
%%%%%%%%% Description of the section %%%%%%%%%%%%%%%%%
% For Existing scenarios: Pathloss, los probability, material penetration loss model, channel paramters LSP, SSP and number of clusters, rays/cluster related discussion, Absolute Time of Arrival$^{*}$ modeling for InH, UMi, UMa, RMa, and SMa,
% Lead: Hitesh
%%%%%%%%%%%%%%%%%%%%%%%%%%%%%%%%%%%%%%%%%%%%%%%%%%%%%%%
Several channel model parameters were updated and newly introduced in 3GPP Rel-19 SI as shown in Table \ref{tab:updates}, which are as follows:
\begin{itemize}
    \item Based on the data provided in Table \ref{tab:meas}, it was identified that DS, ASD, ASA, and ZSA for UMi and UMa scenarios required updates. These updated parameters are captured in Table 7.5.6 Part 1 \cite{tr38901v19} and were derived from both legacy datasets used in 3GPP Rel-14 SI \cite{5g-workshop} and newly acquired data from 3GPP Rel-19 SI using either weighted least squares curve fitting or weighted mean.
    \item Further analysis of both legacy measurements \cite{5g-workshop} and new data provided in 3GPP Rel-19 SI confirmed that there is no strong frequency dependency in determining the number of clusters and showed that the number of clusters per BS-UT link may not be fixed in each scenario as stated in Tables 7.5.6's \cite{tr38901}. This is due to significant variations in real-world deployments depending on multiple factors such as the propagation scenario, carrier frequency, system bandwidth, and spatial resolution. Thus, a framework was introduced in Section 7.6.15 \cite{tr38901v19}, where the number of clusters per BS-UT link can be variable and is selected based on deployments from a bounded interval, as defined in Table 7.6.15-1 \cite{tr38901v19}.
    \item The modeling of absolute time of arrival for non-line of sight (NLOS) was extended across all scenarios based on extensive ray tracing simulations to support sensing and positioning use cases.
\end{itemize}
Additionally, channel sparsity increases with frequency, resulting in fewer dominant multipath components and reduced channel rank. However, Table 7.5.6 \cite{tr38901} imposed fixed number of rays per cluster for narrow band systems and Eq. (7.6-8) \cite{tr38901} imposed a minimum of 20 rays per cluster for ELAAs and/or wideband system across all frequency bands and scenarios. This fixed and minimum lower bound for the number of intra cluster rays did not accommodate the presence of fewer dominant multipath components as observed in measurements. To address this, Eq. (7.6-8) \cite{tr38901v19} was updated to allow modeling of a variable number of rays per cluster for ELAAs and/or wideband system, which can be less than 20 and frequency dependent for various scenarios.
\renewcommand{\arraystretch}{1.2}
\begin{table*}
  \centering
  \caption{Updated and Newly Introduced (*) Channel Model Parameters in 3GPP Rel-19 SI for various scenarios, applicable over the entire frequency range of 0.5–100 GHz. ``NA" indicates that the original values are retained.}
    \begin{tabular}{ccccccccc}
     &
       &
       &
       &
       &
       &
       &
       &
      
      \\
\cline{2-8}    \multicolumn{1}{c|}{} &
      \multicolumn{1}{c|}{\textbf{Scenario}} &
      \multicolumn{3}{c|}{\textbf{Parameters}} &
      \multicolumn{1}{c|}{\textbf{LOS}} &
      \multicolumn{1}{c|}{\textbf{NLOS}} &
      \multicolumn{1}{c|}{\textbf{O2I}} &
      
      \\
\cline{2-8}    \multicolumn{1}{c|}{} &
      \multicolumn{1}{c|}{\multirow{22}{*}{\parbox{4em}{\centering UMi}}} &
      \multicolumn{1}{c|}{\multirow{4}{*}{\parbox{9em}{\centering Delay Spread (DS) lgDS = log$_{10}$(DS/1s)}}} &
      \multicolumn{1}{c|}{\multirow{2}{*}{original}} &
      \multicolumn{1}{c|}{$\mu_{lgDS}$} &
      \multicolumn{1}{c|}{-0.24 log$_{10}$(1+$f_{c}$) - 7.14} &
      \multicolumn{1}{c|}{-0.24 log$_{10}$(1+$f_{c}$) -6.83} &
      \multicolumn{1}{c|}{-6.62} &
      
      \\
\cline{5-8}    \multicolumn{1}{c|}{} &
      \multicolumn{1}{c|}{} &
      \multicolumn{1}{c|}{} &
      \multicolumn{1}{c|}{} &
      \multicolumn{1}{c|}{$\sigma_{lgDS}$} &
      \multicolumn{1}{c|}{0.38} &
      \multicolumn{1}{c|}{0.16 log$_{10}$(1+$f_{c}$) + 0.28} &
      \multicolumn{1}{c|}{0.32} &
      
      \\
\cline{4-8}    \multicolumn{1}{c|}{} &
      \multicolumn{1}{c|}{} &
      \multicolumn{1}{c|}{} &
      \multicolumn{1}{c|}{\multirow{2}{*}{updated}} &
      \multicolumn{1}{c|}{$\mu_{lgDS}$} &
      \multicolumn{1}{c|}{-0.18 log$_{10}$(1+$f_{c}$) - 7.28} &
      \multicolumn{1}{c|}{-0.22 log$_{10}$(1+$f_{c}$) - 6.87} &
      \multicolumn{1}{c|}{\multirow{2}{*}{NA}} &
      
      \\
\cline{5-7}    \multicolumn{1}{c|}{} &
      \multicolumn{1}{c|}{} &
      \multicolumn{1}{c|}{} &
      \multicolumn{1}{c|}{} &
      \multicolumn{1}{c|}{$\sigma_{lgDS}$} &
      \multicolumn{1}{c|}{0.39} &
      \multicolumn{1}{c|}{0.19 log$_{10}$(1+$f_{c}$) + 0.22} &
      \multicolumn{1}{c|}{} &
      
      \\
\cline{3-8}    \multicolumn{1}{c|}{} &
      \multicolumn{1}{c|}{} &
      \multicolumn{1}{c|}{\multirow{4}{*}{\parbox{11em}{\centering AOD Spread (ASD) lgASD = log$_{10}$(ASD/1$^\circ$)}}} &
      \multicolumn{1}{c|}{\multirow{2}{*}{original}} &
      \multicolumn{1}{c|}{$\mu_{lgASD}$} &
      \multicolumn{1}{c|}{-0.05 log$_{10}$(1+$f_{c}$) + 1.21} &
      \multicolumn{1}{c|}{-0.23 log$_{10}$(1+$f_{c}$) + 1.53} &
      \multicolumn{1}{c|}{1.25} &
      
      \\
\cline{5-8}    \multicolumn{1}{c|}{} &
      \multicolumn{1}{c|}{} &
      \multicolumn{1}{c|}{} &
      \multicolumn{1}{c|}{} &
      \multicolumn{1}{c|}{$\sigma_{lgASD}$} &
      \multicolumn{1}{c|}{0.41} &
      \multicolumn{1}{c|}{0.11 log$_{10}$(1+$f_{c}$) + 0.33} &
      \multicolumn{1}{c|}{0.42} &
      
      \\
\cline{4-8}    \multicolumn{1}{c|}{} &
      \multicolumn{1}{c|}{} &
      \multicolumn{1}{c|}{} &
      \multicolumn{1}{c|}{\multirow{2}{*}{updated}} &
      \multicolumn{1}{c|}{$\mu_{lgASD}$} &
      \multicolumn{1}{c|}{-0.05 log$_{10}$(1+$f_{c}$) + 1.21} &
      \multicolumn{1}{c|}{
       -0.24 log$_{10}$(1+$f_{c}$) + 1.54
      } &
      \multicolumn{1}{c|}{\multirow{2}{*}{NA}} &
      
      \\
\cline{5-7}    \multicolumn{1}{c|}{} &
      \multicolumn{1}{c|}{} &
      \multicolumn{1}{c|}{} &
      \multicolumn{1}{c|}{} &
      \multicolumn{1}{c|}{$\sigma_{lgASD}$} &
      \multicolumn{1}{c|}{0.08 log$_{10}$(1+$f_{c}$) + 0.29} &
      \multicolumn{1}{c|}{0.10 log$_{10}$(1+$f_{c}$) + 0.33} &
      \multicolumn{1}{c|}{} &
      
      \\
\cline{3-8}    \multicolumn{1}{c|}{} &
      \multicolumn{1}{c|}{} &
       \multicolumn{1}{c|}{\multirow{4}{*}{\parbox{11em}{\centering AOA Spread (ASA) lgASA = log$_{10}$(ASA/1$^\circ$)}}} &
      \multicolumn{1}{c|}{\multirow{2}{*}{original}} &
      \multicolumn{1}{c|}{$\mu_{lgASA}$} &
      \multicolumn{1}{c|}{-0.08 log$_{10}$(1+$f_{c}$) + 1.73} &
      \multicolumn{1}{c|}{-0.08 log$_{10}$(1+$f_{c}$) + 1.81} &
      \multicolumn{1}{c|}{1.76} &
      
      \\
\cline{5-8}    \multicolumn{1}{c|}{} &
      \multicolumn{1}{c|}{} &
      \multicolumn{1}{c|}{} &
      \multicolumn{1}{c|}{} &
      \multicolumn{1}{c|}{$\sigma_{lgASA}$} &
      \multicolumn{1}{c|}{0.014 log$_{10}$(1+$f_{c}$) + 0.28} &
      \multicolumn{1}{c|}{0.05 log$_{10}$(1+$f_{c}$) + 0.3} &
      \multicolumn{1}{c|}{0.16} &
      
      \\
\cline{4-8}    \multicolumn{1}{c|}{} &
      \multicolumn{1}{c|}{} &
      \multicolumn{1}{c|}{} &
      \multicolumn{1}{c|}{\multirow{2}{*}{updated}} &
      \multicolumn{1}{c|}{$\mu_{lgASA}$} &
      \multicolumn{1}{c|}{-0.07 log$_{10}$(1+$f_{c}$) + 1.66} &
      \multicolumn{1}{c|}{-0.07 log$_{10}$(1+$f_{c}$) + 1.76} &
      \multicolumn{1}{c|}{\multirow{2}{*}{NA}} &
      
      \\
\cline{5-7}    \multicolumn{1}{c|}{} &
      \multicolumn{1}{c|}{} &
      \multicolumn{1}{c|}{} &
      \multicolumn{1}{c|}{} &
      \multicolumn{1}{c|}{$\sigma_{lgASA}$} &
      \multicolumn{1}{c|}{0.021 log$_{10}$(1+$f_{c}$) + 0.26} &
      \multicolumn{1}{c|}{0.05 log$_{10}$(1+$f_{c}$) + 0.27} &
      \multicolumn{1}{c|}{} &
      
      \\
\cline{3-8}    \multicolumn{1}{c|}{} &
      \multicolumn{1}{c|}{} &
      \multicolumn{1}{c|}{\multirow{4}{*}{\parbox{11em}{\centering ZOA Spread (ZSA) lgZSA = log$_{10}$(ZSA/1$^\circ$)}}} &
      \multicolumn{1}{c|}{\multirow{2}{*}{original}} &
      \multicolumn{1}{c|}{$\mu_{lgZSA}$} &
      \multicolumn{1}{c|}{-0.1 log$_{10}$(1+$f_{c}$) + 0.73} &
      \multicolumn{1}{c|}{-0.04 log$_{10}$(1+$f_{c}$) + 0.92} &
      \multicolumn{1}{c|}{1.01} &
      
      \\
\cline{5-8}    \multicolumn{1}{c|}{} &
      \multicolumn{1}{c|}{} &
      \multicolumn{1}{c|}{} &
      \multicolumn{1}{c|}{} &
      \multicolumn{1}{c|}{$\sigma_{lgZSA}$} &
      \multicolumn{1}{c|}{-0.04 log$_{10}$(1+$f_{c}$) + 0.34} &
      \multicolumn{1}{c|}{-0.07 log$_{10}$(1+$f_{c}$) + 0.41} &
      \multicolumn{1}{c|}{0.43} &
      
      \\
\cline{4-8}    \multicolumn{1}{c|}{} &
      \multicolumn{1}{c|}{} &
      \multicolumn{1}{c|}{} &
      \multicolumn{1}{c|}{\multirow{2}{*}{updated}} &
      \multicolumn{1}{c|}{$\mu_{lgZSA}$} &
      \multicolumn{1}{c|}{-0.11 log$_{10}$(1+$f_{c}$) + 0.81} &
      \multicolumn{1}{c|}{-0.03 log$_{10}$(1+$f_{c}$) + 0.92} &
      \multicolumn{1}{c|}{\multirow{2}{*}{NA}} &
      
      \\
\cline{5-7}    \multicolumn{1}{c|}{} &
      \multicolumn{1}{c|}{} &
      \multicolumn{1}{c|}{} &
      \multicolumn{1}{c|}{} &
      \multicolumn{1}{c|}{$\sigma_{lgZSA}$} &
      \multicolumn{1}{c|}{-0.03 log$_{10}$(1+$f_{c}$) + 0.29} &
      \multicolumn{1}{c|}{-0.05 log$_{10}$(1+$f_{c}$) + 0.35} &
      \multicolumn{1}{c|}{} &
      
      \\
\cline{3-8}    \multicolumn{1}{c|}{} &
      \multicolumn{1}{c|}{} &
      \multicolumn{1}{c|}{\multirow{3}{*}{Number of Clusters \textit{N}}} &
      \multicolumn{2}{c|}{original} &
      \multicolumn{1}{c|}{12} &
      \multicolumn{1}{c|}{19} &
      \multicolumn{1}{c|}{12} &
      
      \\
\cline{4-8}    \multicolumn{1}{c|}{} &
      \multicolumn{1}{c|}{} &
      \multicolumn{1}{c|}{} &
      \multicolumn{1}{c|}{\multirow{2}{*}{optional$^{*}$}} &
      \multicolumn{1}{c|}{D$_{min}$} &
      \multicolumn{1}{c|}{6} &
      \multicolumn{1}{c|}{6} &
      \multicolumn{1}{c|}{6} &
      
      \\
\cline{5-8}    \multicolumn{1}{c|}{} &
      \multicolumn{1}{c|}{} &
      \multicolumn{1}{c|}{} &
      \multicolumn{1}{c|}{} &
      \multicolumn{1}{c|}{D$_{max}$} &
      \multicolumn{1}{c|}{12} &
      \multicolumn{1}{c|}{19} &
      \multicolumn{1}{c|}{12} &
      
      \\
\cline{3-8}    \multicolumn{1}{c|}{} &
      \multicolumn{1}{c|}{} &
      \multicolumn{1}{c|}{\multirow{3}[6]{*}{Absolute Time of Arrival$^{*}$}} &
      \multicolumn{1}{c|}{\multirow{2}{*}{\parbox{5.5em}{\centering $lg\Delta\tau = log_{10}(\Delta\tau/1s)$}}} &
      \multicolumn{1}{c|}{$\mu_{lg\Delta\tau}$} &
      \multicolumn{1}{c|}{\multirow{3}{*}{NA}} &
      \multicolumn{1}{c|}{-7.5} &
      \multicolumn{1}{c|}{\multirow{3}{*}{NA}} &
      
      \\
\cline{5-5}\cline{7-7}    \multicolumn{1}{c|}{} &
      \multicolumn{1}{c|}{} &
      \multicolumn{1}{c|}{} &
      \multicolumn{1}{c|}{} &
      \multicolumn{1}{c|}{$\sigma_{lg\Delta\tau}$} &
      \multicolumn{1}{c|}{} &
      \multicolumn{1}{c|}{0.5} &
      \multicolumn{1}{c|}{} &
      
      \\
\cline{4-5}\cline{7-7}    \multicolumn{1}{c|}{} &
      \multicolumn{1}{c|}{} &
      \multicolumn{1}{c|}{} &
      \multicolumn{2}{p{10em}|}{Correlation distance in the horizontal plane [m]} &
      \multicolumn{1}{c|}{} &
      \multicolumn{1}{c|}{15} &
      \multicolumn{1}{c|}{} &
      
      \\
\cline{2-8}    \multicolumn{1}{c|}{} &
      \multicolumn{1}{c|}{\multirow{24}{*}{\parbox{5em}{\centering UMa}}} &
      \multicolumn{1}{c|}{\multirow{4}{*}{\parbox{9em}{\centering Delay Spread (DS) lgDS = log$_{10}$(DS/1s)}}} &
      \multicolumn{1}{c|}{\multirow{2}{*}{original}} &
      \multicolumn{1}{c|}{$\mu_{lgDS}$} &
      \multicolumn{1}{c|}{-6.955 - 0.0963log$_{10}$($f_{c}$)} &
      \multicolumn{1}{c|}{-6.28 - 0.204 log$_{10}$($f_{c}$)} &
      \multicolumn{1}{c|}{-6.62} &
      
      \\
\cline{5-8}    \multicolumn{1}{c|}{} &
      \multicolumn{1}{c|}{} &
      \multicolumn{1}{c|}{} &
      \multicolumn{1}{c|}{} &
      \multicolumn{1}{c|}{$\sigma_{lgDS}$} &
      \multicolumn{1}{c|}{0.66} &
      \multicolumn{1}{c|}{0.39} &
      \multicolumn{1}{c|}{0.32} &
      
      \\
\cline{4-8}    \multicolumn{1}{c|}{} &
      \multicolumn{1}{c|}{} &
      \multicolumn{1}{c|}{} &
      \multicolumn{1}{c|}{\multirow{2}{*}{updated}} &
      \multicolumn{1}{c|}{$\mu_{lgDS}$} &
      \multicolumn{1}{c|}{-7.067 - 0.0794 log$_{10}$($f_{c}$)} &
      \multicolumn{1}{c|}{-6.47 - 0.134 log$_{10}$($f_{c}$)} &
      \multicolumn{1}{c|}{\multirow{2}{*}{NA}} &
      
      \\
\cline{5-7}    \multicolumn{1}{c|}{} &
      \multicolumn{1}{c|}{} &
      \multicolumn{1}{c|}{} &
      \multicolumn{1}{c|}{} &
      \multicolumn{1}{c|}{$\sigma_{lgDS}$} &
      \multicolumn{1}{c|}{0.57 + 0.026 log$_{10}$($f_{c}$)} &
      \multicolumn{1}{c|}{0.39} &
      \multicolumn{1}{c|}{} &
      
      \\
\cline{3-8}    \multicolumn{1}{c|}{} &
      \multicolumn{1}{c|}{} &
      \multicolumn{1}{c|}{\multirow{4}{*}{\parbox{11em}{\centering AOD Spread (ASD) lgASD = log$_{10}$(ASD/1$^\circ$)}}} &
      \multicolumn{1}{c|}{\multirow{2}{*}{original}} &
      \multicolumn{1}{c|}{$\mu_{lgASD}$} &
      \multicolumn{1}{c|}{1.06 + 0.1114 log$_{10}$($f_{c}$)} &
      \multicolumn{1}{c|}{1.5 - 0.1144 log$_{10}$($f_{c}$)} &
      \multicolumn{1}{c|}{1.25} &
      
      \\
\cline{5-8}    \multicolumn{1}{c|}{} &
      \multicolumn{1}{c|}{} &
      \multicolumn{1}{c|}{} &
      \multicolumn{1}{c|}{} &
      \multicolumn{1}{c|}{$\sigma_{lgASD}$} &
      \multicolumn{1}{c|}{0.28} &
      \multicolumn{1}{c|}{0.28} &
      \multicolumn{1}{c|}{0.42} &
      
      \\
\cline{4-8}    \multicolumn{1}{c|}{} &
      \multicolumn{1}{c|}{} &
      \multicolumn{1}{c|}{} &
      \multicolumn{1}{c|}{\multirow{2}{*}{updated}} &
      \multicolumn{1}{c|}{$\mu_{lgASD}$} &
      \multicolumn{1}{c|}{0.92} &
      \multicolumn{1}{c|}{1.09} &
      \multicolumn{1}{c|}{0.58} &
      
      \\
\cline{5-8}    \multicolumn{1}{c|}{} &
      \multicolumn{1}{c|}{} &
      \multicolumn{1}{c|}{} &
      \multicolumn{1}{c|}{} &
      \multicolumn{1}{c|}{$\sigma_{lgASD}$} &
      \multicolumn{1}{c|}{0.31} &
      \multicolumn{1}{c|}{0.44} &
      \multicolumn{1}{c|}{0.7} &
      
      \\
\cline{3-8}    \multicolumn{1}{c|}{} &
      \multicolumn{1}{c|}{} &
       \multicolumn{1}{c|}{\multirow{4}{*}{\parbox{11em}{\centering AOA Spread (ASA) lgASA = log$_{10}$(ASA/1$^\circ$)}}} &
      \multicolumn{1}{c|}{\multirow{2}{*}{original}} &
      \multicolumn{1}{c|}{$\mu_{lgASA}$} &
      \multicolumn{1}{c|}{1.81} &
      \multicolumn{1}{c|}{2.08 - 0.27 log$_{10}$($f_{c}$)} &
      \multicolumn{1}{c|}{1.76} &
      
      \\
\cline{5-8}    \multicolumn{1}{c|}{} &
      \multicolumn{1}{c|}{} &
      \multicolumn{1}{c|}{} &
      \multicolumn{1}{c|}{} &
      \multicolumn{1}{c|}{$\sigma_{lgASA}$} &
      \multicolumn{1}{c|}{0.2} &
      \multicolumn{1}{c|}{0.11} &
      \multicolumn{1}{c|}{0.16} &
      
      \\
\cline{4-8}    \multicolumn{1}{c|}{} &
      \multicolumn{1}{c|}{} &
      \multicolumn{1}{c|}{} &
      \multicolumn{1}{c|}{\multirow{2}{*}{updated}} &
      \multicolumn{1}{c|}{$\mu_{lgASA}$} &
      \multicolumn{1}{c|}{1.76} &
      \multicolumn{1}{c|}{2.04 - 0.25 log$_{10}$($f_{c}$)} &
      \multicolumn{1}{c|}{\multirow{2}{*}{NA}} &
      
      \\
\cline{5-7}    \multicolumn{1}{c|}{} &
      \multicolumn{1}{c|}{} &
      \multicolumn{1}{c|}{} &
      \multicolumn{1}{c|}{} &
      \multicolumn{1}{c|}{$\sigma_{lgASA}$} &
      \multicolumn{1}{c|}{0.19} &
      \multicolumn{1}{c|}{0.17 - 0.03 log$_{10}$($f_{c}$)} &
      \multicolumn{1}{c|}{} &
      
      \\
\cline{3-8}    \multicolumn{1}{c|}{} &
      \multicolumn{1}{c|}{} &
      \multicolumn{1}{c|}{\multirow{4}{*}{\parbox{11em}{\centering ZOA Spread (ZSA) lgZSA = log$_{10}$(ZSA/1$^\circ$)}}} &
      \multicolumn{1}{c|}{\multirow{2}{*}{original}} &
      \multicolumn{1}{c|}{$\mu_{lgZSA}$} &
      \multicolumn{1}{c|}{0.95} &
      \multicolumn{1}{c|}{1.512 - 0.3236 log$_{10}$($f_{c}$)} &
      \multicolumn{1}{c|}{1.01} &
      
      \\
\cline{5-8}    \multicolumn{1}{c|}{} &
      \multicolumn{1}{c|}{} &
      \multicolumn{1}{c|}{} &
      \multicolumn{1}{c|}{} &
      \multicolumn{1}{c|}{$\sigma_{lgZSA}$} &
      \multicolumn{1}{c|}{0.16} &
      \multicolumn{1}{c|}{0.16} &
      \multicolumn{1}{c|}{0.43} &
      
      \\
\cline{4-8}    \multicolumn{1}{c|}{} &
      \multicolumn{1}{c|}{} &
      \multicolumn{1}{c|}{} &
      \multicolumn{1}{c|}{\multirow{2}{*}{updated}} &
      \multicolumn{1}{c|}{$\mu_{lgZSA}$} &
      \multicolumn{1}{c|}{0.96} &
      \multicolumn{1}{c|}{1.445 - 0.2856 log$_{10}$($f_{c}$)} &
      \multicolumn{1}{c|}{\multirow{2}{*}{NA}} &
      
      \\
\cline{5-7}    \multicolumn{1}{c|}{} &
      \multicolumn{1}{c|}{} &
      \multicolumn{1}{c|}{} &
      \multicolumn{1}{c|}{} &
      \multicolumn{1}{c|}{$\sigma_{lgZSA}$} &
      \multicolumn{1}{c|}{0.15} &
      \multicolumn{1}{c|}{0.17} &
      \multicolumn{1}{c|}{} &
      
      \\
\cline{3-8}    \multicolumn{1}{c|}{} &
      \multicolumn{1}{c|}{} &
      \multicolumn{1}{c|}{\multirow{3}{*}{Number of Clusters \textit{N}}} &
      \multicolumn{2}{c|}{original} &
      \multicolumn{1}{c|}{12} &
      \multicolumn{1}{c|}{20} &
      \multicolumn{1}{c|}{12} &
      
      \\
\cline{4-8}    \multicolumn{1}{c|}{} &
      \multicolumn{1}{c|}{} &
      \multicolumn{1}{c|}{} &
      \multicolumn{1}{c|}{\multirow{2}{*}{optional$^{*}$}} &
      \multicolumn{1}{c|}{D$_{min}$} &
      \multicolumn{1}{c|}{10} &
      \multicolumn{1}{c|}{15} &
      \multicolumn{1}{c|}{10} &
      
      \\
\cline{5-8}    \multicolumn{1}{c|}{} &
      \multicolumn{1}{c|}{} &
      \multicolumn{1}{c|}{} &
      \multicolumn{1}{c|}{} &
      \multicolumn{1}{c|}{D$_{max}$} &
      \multicolumn{1}{c|}{12} &
      \multicolumn{1}{c|}{20} &
      \multicolumn{1}{c|}{12} &
      
      \\
\cline{3-8}    \multicolumn{1}{c|}{} &
      \multicolumn{1}{c|}{} &
      \multicolumn{1}{c|}{\multirow{2}{*}{Cluster ASD (c$_{ASD}$) in [deg]}} &
      \multicolumn{2}{c|}{original} &
      \multicolumn{1}{c|}{5} &
      \multicolumn{1}{c|}{2} &
      \multicolumn{1}{c|}{5} &
      
      \\
\cline{4-8}    \multicolumn{1}{c|}{} &
      \multicolumn{1}{c|}{} &
      \multicolumn{1}{c|}{} &
      \multicolumn{2}{c|}{updated} &
      \multicolumn{1}{c|}{3.58} &
      \multicolumn{1}{c|}{1.8} &
      \multicolumn{1}{c|}{1.8} &
      
      \\
\cline{3-8}    \multicolumn{1}{c|}{} &
      \multicolumn{1}{c|}{} &
      \multicolumn{1}{c|}{\multirow{3}[6]{*}{Absolute Time of Arrival$^{*}$}} &
      \multicolumn{1}{c|}{\multirow{2}{*}{\parbox{5.5em}{\centering $lg\Delta\tau = log_{10}(\Delta\tau/1s)$}}} &
      \multicolumn{1}{c|}{$\mu_{lg\Delta\tau}$} &
      \multicolumn{1}{c|}{\multirow{3}{*}{NA}} &
      \multicolumn{1}{c|}{-7.4} &
      \multicolumn{1}{c|}{\multirow{3}{*}{NA}} &
      
      \\
\cline{5-5}\cline{7-7}    \multicolumn{1}{c|}{} &
      \multicolumn{1}{c|}{} &
      \multicolumn{1}{c|}{} &
      \multicolumn{1}{c|}{} &
      \multicolumn{1}{c|}{$\sigma_{lg\Delta\tau}$} &
      \multicolumn{1}{c|}{} &
      \multicolumn{1}{c|}{0.2} &
      \multicolumn{1}{c|}{} &
      
      \\
\cline{4-5}\cline{7-7}    \multicolumn{1}{c|}{} &
      \multicolumn{1}{c|}{} &
      \multicolumn{1}{c|}{} &
      \multicolumn{2}{p{10em}|}{Correlation distance in the horizontal plane [m]} &
      \multicolumn{1}{c|}{} &
      \multicolumn{1}{c|}{50} &
      \multicolumn{1}{c|}{} &
      
      \\
\cline{2-8}    \multicolumn{1}{c|}{} &
      \multicolumn{1}{c|}{\multirow{6}{*}{\parbox{4em}{\centering InH}}} &
      \multicolumn{1}{c|}{\multirow{3}{*}{Number of Clusters \textit{N}}} &
      \multicolumn{2}{c|}{original} &
      \multicolumn{1}{c|}{15} &
      \multicolumn{1}{c|}{19} &
      \multicolumn{1}{c|}{\multirow{6}[12]{*}{NA}} &
      
      \\
\cline{4-7}    \multicolumn{1}{c|}{} &
      \multicolumn{1}{c|}{} &
      \multicolumn{1}{c|}{} &
      \multicolumn{1}{c|}{\multirow{2}{*}{optional$^{*}$}} &
      \multicolumn{1}{c|}{D$_{min}$} &
      \multicolumn{1}{c|}{7} &
      \multicolumn{1}{c|}{6} &
      \multicolumn{1}{c|}{} &
      
      \\
\cline{5-7}    \multicolumn{1}{c|}{} &
      \multicolumn{1}{c|}{} &
      \multicolumn{1}{c|}{} &
      \multicolumn{1}{c|}{} &
      \multicolumn{1}{c|}{D$_{max}$} &
      \multicolumn{1}{c|}{15} &
      \multicolumn{1}{c|}{19} &
      \multicolumn{1}{c|}{} &
      
      \\
\cline{3-7}    \multicolumn{1}{c|}{} &
      \multicolumn{1}{c|}{} &
      \multicolumn{1}{c|}{\multirow{3}[6]{*}{Absolute Time of Arrival$^{*}$}} &
    \multicolumn{1}{c|}{\multirow{2}{*}{\parbox{5.5em}{\centering $lg\Delta\tau = log_{10}(\Delta\tau/1s)$}}} &
      \multicolumn{1}{c|}{$\mu_{lg
      \Delta\tau}$} &
      \multicolumn{1}{c|}{\multirow{3}{*}{NA}} &
      \multicolumn{1}{c|}{-8.6} &
      \multicolumn{1}{c|}{} &
      
      \\
\cline{5-5}\cline{7-7}    \multicolumn{1}{c|}{} &
      \multicolumn{1}{c|}{} &
      \multicolumn{1}{c|}{} &
      \multicolumn{1}{c|}{} &
      \multicolumn{1}{c|}{$\sigma_{lg\Delta\tau}$} &
      \multicolumn{1}{c|}{} &
      \multicolumn{1}{c|}{0.1} &
      \multicolumn{1}{c|}{} &
      
      \\
\cline{4-5}\cline{7-7}    \multicolumn{1}{c|}{} &
      \multicolumn{1}{c|}{} &
      \multicolumn{1}{c|}{} &
      \multicolumn{2}{p{10em}|}{Correlation distance in the horizontal plane [m]} &
      \multicolumn{1}{c|}{} &
      \multicolumn{1}{c|}{10} &
      \multicolumn{1}{c|}{} &
      
      \\
\cline{2-8}    \multicolumn{1}{c|}{} &
      \multicolumn{1}{c|}{\multirow{3}{*}{\parbox{5em}{\vspace{0.8em}\centering RMa}}} &
      \multicolumn{1}{c|}{\multirow{3}[6]{*}{Absolute Time of Arrival$^{*}$}} &
     \multicolumn{1}{c|}{\multirow{2}{*}{\parbox{5.5em}{\centering $lg\Delta\tau = log_{10}(\Delta\tau/1s)$}}} &
      \multicolumn{1}{c|}{$\mu_{lg\Delta\tau}$} &
      \multicolumn{1}{c|}{\multirow{3}{*}{NA}} &
      \multicolumn{1}{c|}{-8.33} &
      \multicolumn{1}{c|}{\multirow{3}{*}{NA}} &
      
      \\
\cline{5-5}\cline{7-7}    \multicolumn{1}{c|}{} &
      \multicolumn{1}{c|}{} &
      \multicolumn{1}{c|}{} &
      \multicolumn{1}{c|}{} &
      \multicolumn{1}{c|}{$\sigma_{lg\Delta\tau}$} &
      \multicolumn{1}{c|}{} &
      \multicolumn{1}{c|}{0.26} &
      \multicolumn{1}{c|}{} &
      
      \\
\cline{4-5}\cline{7-7}    \multicolumn{1}{c|}{} &
      \multicolumn{1}{c|}{} &
      \multicolumn{1}{c|}{} &
      \multicolumn{2}{p{10em}|}{Correlation distance in the horizontal plane [m]} &
      \multicolumn{1}{c|}{} &
      \multicolumn{1}{c|}{50} &
      \multicolumn{1}{c|}{} &
      
      \\
\cline{2-8}     &
       &
       &
       &
       &
       &
       &
       &
      
      \\
    \end{tabular}%
  \label{tab:updates}%
\end{table*}%
\renewcommand{\arraystretch}{1}

\subsection{Polarization}\label{sec:polarization}
%%%%%%%%% Description of the section %%%%%%%%%%%%%%%%%
% dicussions related to introduction of pol variability
% Lead: Henrick
%%%%%%%%%%%%%%%%%%%%%%%%%%%%%%%%%%%%%%%%%%%%%%%%%%%%%%%
The use of dual-polarized antennas can provide significant benefits in diversity and MIMO multiplexing gains. % Hence, they have become a de facto standard in BS antenna design, including large antenna arrays for massive MIMO systems. Similarly, UT antenna design also strives for utilizing polarization diversity, to the extent that is feasible, as discussed in Section \ref{sec:ue_ant_model}. 
A key component in channel modeling is the characterization and representation of polarization transformation i.e., how transmitted waves with certain polarizations are altered through interactions in the propagation environment into possibly different polarizations before arriving at the receiver. This transformation is represented by a 2x2 polarization matrix in 7.5-28 \cite{tr38901}, where each element of the matrix captures the amplitude and phase change corresponding to different combinations of transmit and receive polarizations. The two diagonal elements represent the case in which the transmitted polarization and the received polarization are the same. By convention, these polarizations are either co-polar vertical polarization (VP) or co-polar horizontal polarization (HP). The off-diagonal elements correspond to cross-polar components, where the transmit and receive polarizations are orthogonal, e.g., VP to HP or HP to VP. While any pair of transmit and receive polarizations could be used to define this matrix, the selection of VP and HP both for the transmitter and receiver leads to a diagonally-dominant matrix, i.e. the cross-polar components are generally much weaker in magnitude than the co-polar ones \cite{asplund2007propagation}. It should be noted that this matrix characterizes the polarization transformation due to propagation only; it does not account for the specific polarizations of the antennas themselves. Since real antennas may transmit arbitrary polarizations, their impact must be separately modeled, typically by expressing their radiated fields as linear combinations of VP and HP. When combined with channel's polarization matrix, the effective channel can be evaluated, capturing both antenna and propagation effect.
Prior to the 3GPP Rel-19 SI, the polarization matrix in Eq. (7.5-28) \cite{tr38901} assumed:
\begin{itemize}
    \item The two co-polarized diagonal elements of the polarization matrix had identical magnitudes.
    \item The two cross-polarized off-diagonal elements had identical magnitudes but were weaker than the co-polarized diagonal elements.
    \item All four elements had independent random phases.
\end{itemize}
These assumptions implied a form of symmetry or balance between VP and HP, which influenced several radio interface design choices such as MIMO codebook construction. To validate these assumptions, a thorough re-examination of earlier measurements as well as new measurements of the polarization transformations between the transmit and receive antennas was conducted. While the earlier modeling assumptions  were broadly accurate in terms of identical average powers among co and cross polarization components, with cross polarization components exhibiting weaker power compared to co polar components, the study revealed some variability around these assumed identical magnitudes in real-world propagation environments, which was not captured in the original model in Eq. (7.5-28) \cite{tr38901}. 
To account for the variability in power among the different polarizations, a polarization variability model was introduced in Section 7.6.16 \cite{tr38901v19}. \par In this model, each element of the polarization matrix is multiplied by a random factor that accounts for the variability in power among different polarizations. This factor follows a log normal Gaussian distribution with a standard deviation of 3 dB obtained from measurements \cite{asplund2007propagation}. Furthermore, when this model is used in conjunction with the more realistic UT antenna models described in Section \ref{sec:ue_ant_model}, this polarization variability model can be used to ensure that future 6G systems are robust against realistic polarization properties experienced in real-world scenarios.

\subsection{Near Field Channel Model}\label{sec:nf}
%%%%%%%%% Description of the section %%%%%%%%%%%%%%%%%
% Important discussion points on NF channel model
% Lead: Nan (content provided by Nan over email)
%%%%%%%%%%%%%%%%%%%%%%%%%%%%%%%%%%%%%%%%%%%%%%%%%%%%%%%
The deployment of ELAAs is a key enabler for enhanced mobile broadband, improved spectral efficiency, and increased capacity in next-generation MIMO systems. As the physical aperture of the antenna arrays grows, the propagation transitions from far field to NF. Far field propagation assumes that the wavefront is planar whereas, NF propagation models the wavefront as spherical. The channel models in \cite{tr38901} are based on planar wavefront (far field) and therefore do not accurately capture the characteristics of spherical wavefront (NF). This limitation significantly reduces the applicability of the channel models in \cite{tr38901} for ELAAs deployments, where accurate modeling of NF propagation is essential. Thus, to address this limitation and accurately model NF propagation, several key aspects must be considered:
% \begin{itemize}
%     \item In near field propagation, the channel parameters (e.g., phase and angle) for each cluster vary across each antenna element pair at the BS and UT, resulting in non-linear phase and angle variations.
%     \item Consistency in channel characteristics must be maintained, particularly when a single UT moves or when different UTs are located at various positions. 
% \end{itemize}
\begin{itemize}
\item In NF propagation, the phase and angle for each cluster varies non linearly across each antenna element pair between the BS and UT due to spherical wavefront. This contrasts with far field propagation, where the wavefront are approximated as planar, resulting in linear variations of phase and angle across the antenna array.
\item Consistency in channel characteristics must be maintained, particularly when a single UT moves or when different UTs are located at various positions.
\end{itemize}
A unified channel modeling approach was adopted in 3GPP Rel-19, which captures the new characteristics of NF propagation from the perspective of the antenna array in Section 7.6.13 \cite{tr38901v19} within the framework of the existing far field propagation based stochastic channel models \cite{tr38901}. The core principle in modeling NF propagation within the existing framework \cite{tr38901} involves modeling the antenna element-wise channel parameters between the BS and UT as shown in Fig. \ref{fig:nf_sns} and deriving the corresponding channel coefficients. 
\par For the direct path between the BS and UT, the antenna element-wise phase is calculated based on the LOS distance according to Eq. (7.6-50) \cite{tr38901v19}, while the antenna element-wise angles between the BS and UT are determined based on Eq. (7.6-51) \cite{tr38901v19}. For the non-direct multi-path components, two auxiliary points are introduced to determine the antenna element-wise channel parameters at the BS and UT side as shown in Eq. (7.6-47) and Eq. (7.6-48) \cite{tr38901v19}, respectively. The locations of these points are derived by calculating their distances to the BS or UT based on predefined statistical distributions obtained from measurements according to Table 7.6-13-1 \cite{tr38901v19}. Using the positions of these auxiliary points, along with the locations of the BS and UT antenna elements, the antenna element-wise channel parameters can be computed. Once the antenna element-wise channel parameters (e.g., phase and angle) have been generated, the channel coefficients can be determined by incorporating the effects of the spherical wavefront on both the phase and antenna radiation patterns.
\par % By integrating spherical wavefront and antenna element-wise channel parameters, 
%The NF channel model adopted in Section 7.6.13 \cite{tr38901v19} enables precise performance evaluations and lays a solid foundation for the design and standardization of NF communication systems.

\subsection{Modeling of Spatial Non-Stationarity}\label{sec:sns}
%%%%%%%%% Description of the section %%%%%%%%%%%%%%%%%
% discussions related to SNS
% Lead: Nan (content provided by Nan over email)
%%%%%%%%%%%%%%%%%%%%%%%%%%%%%%%%%%%%%%%%%%%%%%%%%%%%%%%
\begin{figure}[t]
    \centering
    \includegraphics[width=\linewidth]{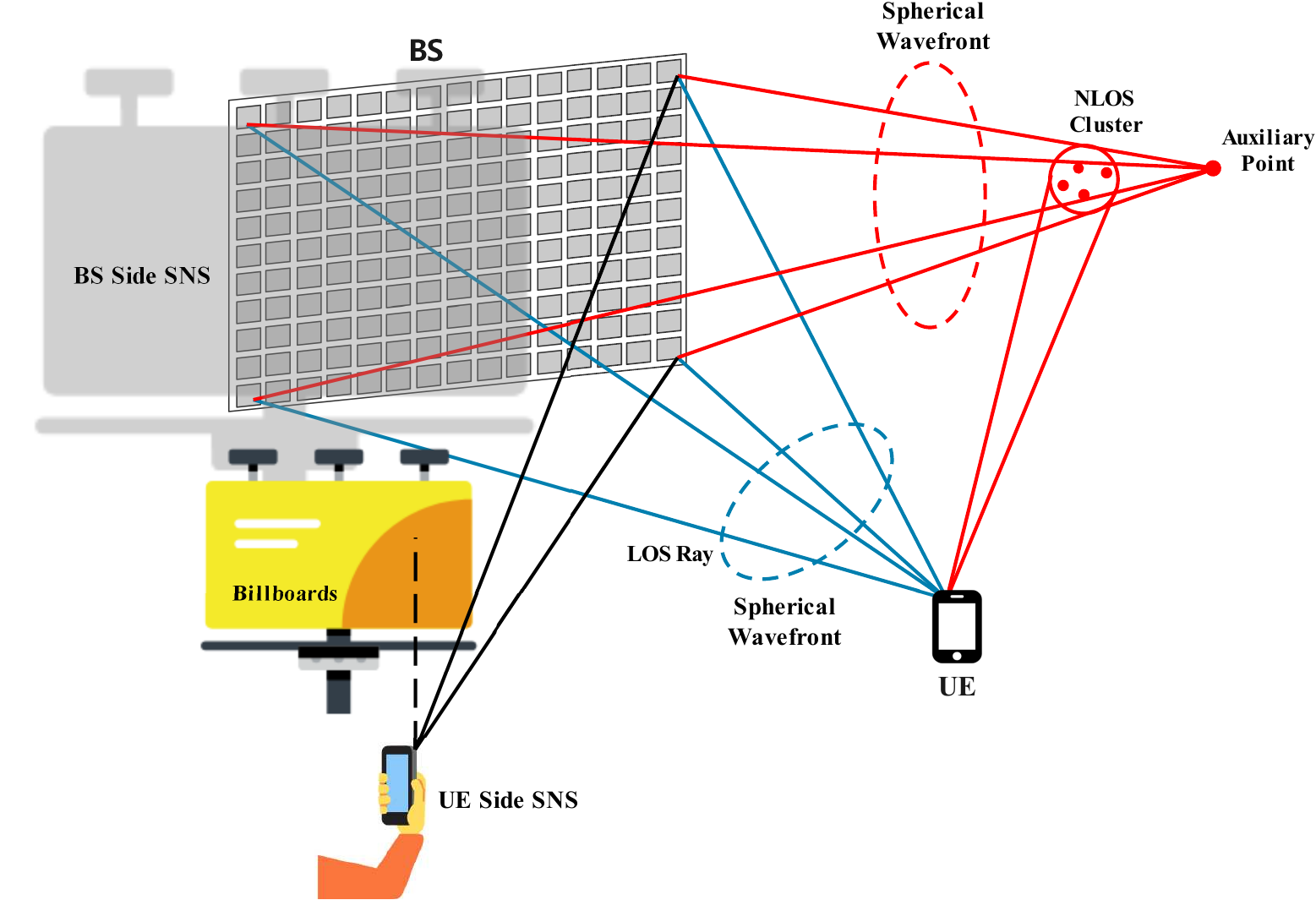}
    \caption{NF propagation and SNS effect at both the BS and UT.}
    \label{fig:nf_sns}
\end{figure}
All the antenna elements within an antenna array at the BS and UT experience similar channel in \cite{tr38901}. However, on deploying ELAAs specifically at the BS, the SNS effects become significant and can no longer be neglected. For instance as shown in Fig. \ref{fig:nf_sns}, physical obstructions near the BS may block only part of the BS antenna array, % and incomplete scattering may prevent some antenna elements from receiving specific clusters. 
and consequently, some clusters are only visible to certain parts of the antenna array, leading to antenna element-wise power variations at the BS. %Under such conditions, the assumptions in TR 38.901 are no longer valid for massive MIMO scenarios, highlighting the need for an accurate SNS model. 
To effectively capture the SNS effects, the following factors must be considered:
\begin{itemize}
    %\item The SNS effect can occur at both the BS and UT. However, it is more pronounced at the BS due to the use of ELAAs, which are typically not feasible within the compact form factor of UTs.
    \item Only a subset of clusters or rays is affected by SNS. Therefore, it is essential to identify which specific clusters or rays are impacted.
    \item Due to SNS, the channel varies across antenna elements. Thus, it is necessary to determine which antenna elements are affected for each cluster or ray. 
    \item The primary effect of SNS is the power variation across the antenna elements. Accurate calculation of antenna element-wise power attenuation factors is therefore required. 
    \item Consistency must be ensured both among clusters and across antenna elements to maintain the reliability and coherence of the SNS channel model.
\end{itemize}
\par For modeling SNS effect at the BS, two alternative approaches are proposed - a physical blocker-based model and a stochastic model. In the physical blocker-based model, the number, locations and dimensions of the blockers are explicitly determined as per Table 7.6.14.1.1-1 \cite{tr38901v19}. Based on the spatial relationship between clusters, blockers, and antenna elements, the antenna element-wise blockage conditions and power attenuation factors are determined based on Eq. (7.6-52) \cite{tr38901v19}. % To represent  typical blockers near the BS, several types of blockers are defined, including billboards, street lamps, building edges and pillars. 
In contrast, the stochastic model uses a statistical approach. In this approach, every UE is assigned a visibility probability that governs the fraction of its clusters impacted by SNS. Next, for each impacted cluster a visibility region is defined that determines the fraction of the BS antennas that are visible to this cluster. The clusters or rays affected by SNS are determined using Eq. (7.6-53, 7.6-54) \cite{tr38901v19}, their corresponding visible regions on the antenna array are computed using Eq. (7.6-57) \cite{tr38901v19}, and the antenna element-wise power attenuation factors in Eq. (7.6-58) \cite{tr38901v19} are randomly generated based on statistical distributions derived from measurement and simulation data. Additionally, the modeling of SNS effect at the UT is described earlier in Section \ref{sec:ue_ant_model}. 
%The modeling of SNS effects, as presented in Section 7.6.14 \cite{tr38901v19}, facilitates the assessment, design, and standardization of next generation systems in which these effects can no longer be neglected.

%%%%% removed this part as per Dimitri's comment - this is already covered in UT antenna modeling section
% \st{For the SNS channel model on the UE side, considering updated UE antenna modeling designs and practical real-world use cases, three typical usage scenarios are taken into account: one-hand grip, dual-hand grip, and head-and-one-hand grip. The element-wise power attenuation is determined based on simulation and measurement results.The proposed SNS channel model provides a flexible framework that supports diverse scenarios and use-case-specific requirements. By incorporating element-wise power variations due to SNS, it enables accurate performance evaluation of massive MIMO systems.}
\section{Conclusion}\label{sec:conclusion}
%%%%%%%%% Description of the section %%%%%%%%%%%%%%%%%
% Conclusing remarks
% Lead: Mansoor, Henrik
%%%%%%%%%%%%%%%%%%%%%%%%%%%%%%%%%%%%%%%%%%%%%%%%%%%%%%%
This paper provides an overview of the 3GPP Rel-19 standardization efforts on channel modeling for 6G. These include the introduction of a new SMa scenario along with its associated modeling components, incorporation of realistic antenna models for handheld UTs, support for modeling variable number of clusters, reduced rays per cluster, and variability in power among different polarizations. Additionally, a new plywood material penetration loss model, and a low loss O2I building penetration loss model for SMa scenario was introduced. Absolute delay modeling was extended to cover scenarios such as UMa, UMi, RMa, InH, and SMa. Moreover, channel parameters such as DS, ASA, ASD, and ZSA for UMi and UMa scenarios were updated based on new measurements and simulations. Also, refinements were made to cluster ASD for UMa LOS, NLOS, and O2I scenarios. Furthermore, channel modeling for NF propagation and SNS effect was developed for evaluating and analyzing future MIMO systems employing ELAA.

% \section*{Appendix}\label{sec:appendix}
% \input{Sections/appendix.tex}
\bibliographystyle{IEEEtran}
% \bibliography{IEEEabrv,ref}

 

% \end{IEEEbiography}


\begin{IEEEbiography}[{\includegraphics[width=1in,height=1.25in,clip,keepaspectratio]{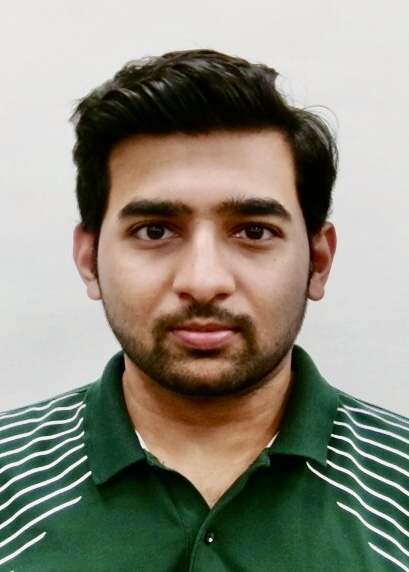}}]
{Hitesh Poddar} (poddarh@sharplabs.com) is a
Senior Communication Systems Researcher at Sharp
Laboratories of America (SLA), Vancouver, Washington, USA and a delegate to 3GPP RAN1. He earned his M.S. in Electrical Engineering from New York University (NYU) in 2023, under the supervision of Prof. Theodore S. Rappaport. His primary research interests include mmWave and sub-THz radio propagation measurement and channel modeling for next generation wireless systems.
\end{IEEEbiography}

\vskip 0pt plus -1fil

\begin{IEEEbiography}
[{\includegraphics[width=1in,height=1.25in,clip,keepaspectratio]{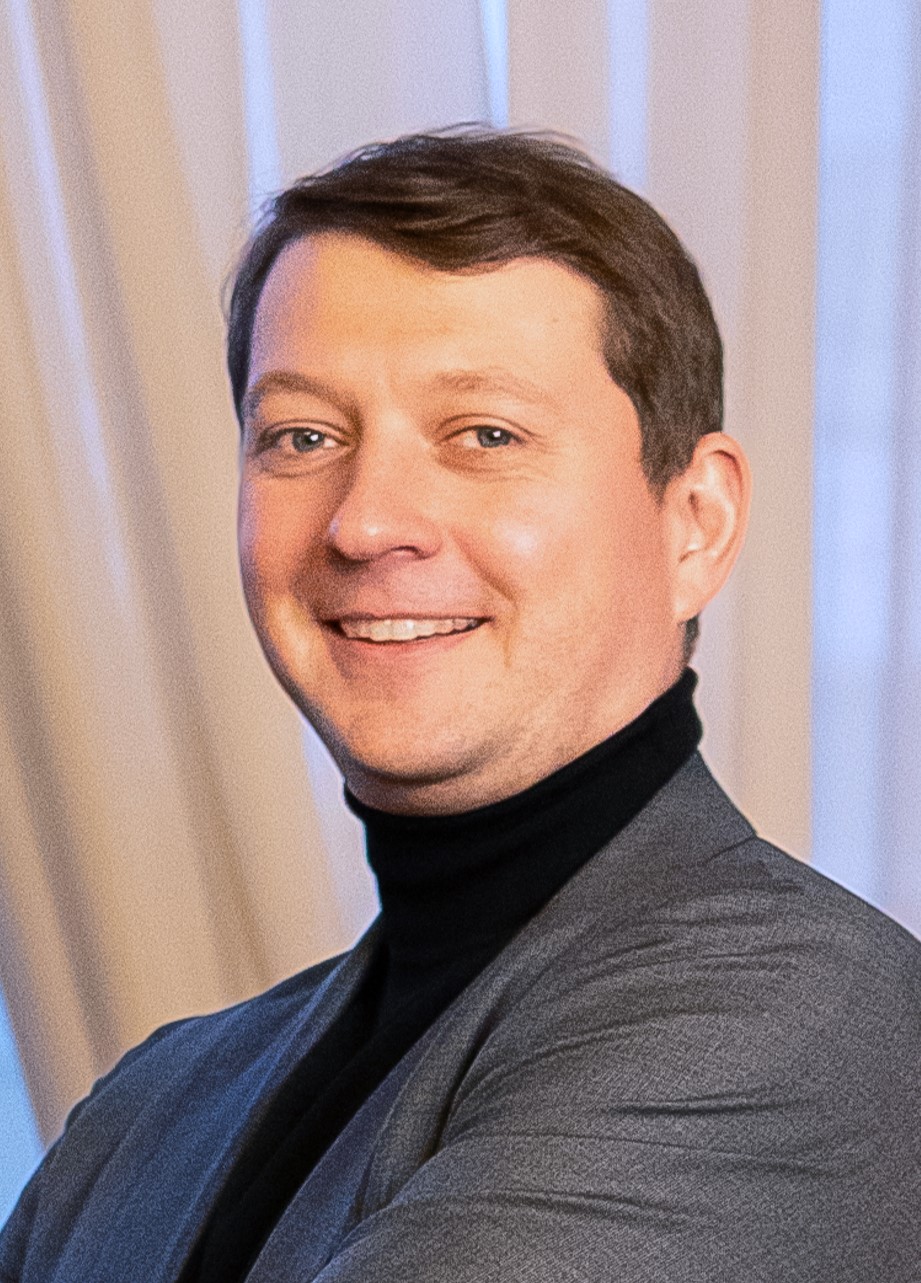}}]
{Dimitri Gold} (dimitri.gold@nokia.com) received the M.Sc. degree in mathematical physics from Moscow State University (2007) and the Ph.D. degree in mathematical information technology and telecommunications from the University of Jyväskylä, Finland (2012), where he also got the title of Docent (2016). He is a 3GPP Standardization delegate at Nokia, focusing on future wireless networks, machine learning in communications, and network modeling.
\end{IEEEbiography}
\vskip 0pt plus -1fil
\begin{IEEEbiography}
[{\includegraphics[width=1in,height=1.25in,clip,keepaspectratio]{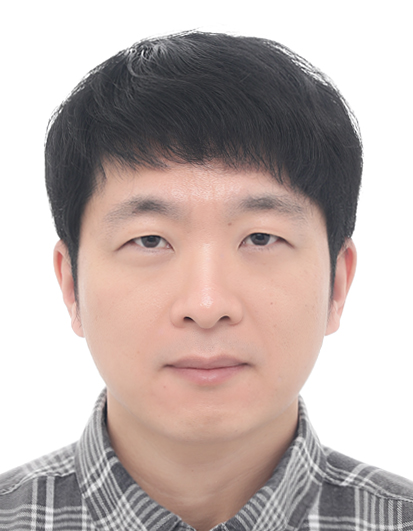}}]
{Daewon Lee} (daewon.lee@intel.com) received Ph.D. degree from Georgia Institute of Technology and B.S. and M.S. degree from Korea Advanced Institute of Science and Technology. He previously worked at LG Electronics and Broadcom for research and development of wireless technology for Wi-Fi and 4G LTE systems. He was also with a startup, Newracom, that developed 802.11 WLAN modems. Since 2016, he has been with Intel Corporation working on 5G NR and 6G technologies.
\end{IEEEbiography}
\vskip 0pt plus -1fil
\begin{IEEEbiography}
[{\includegraphics[width=1in,height=1.25in,clip,keepaspectratio]{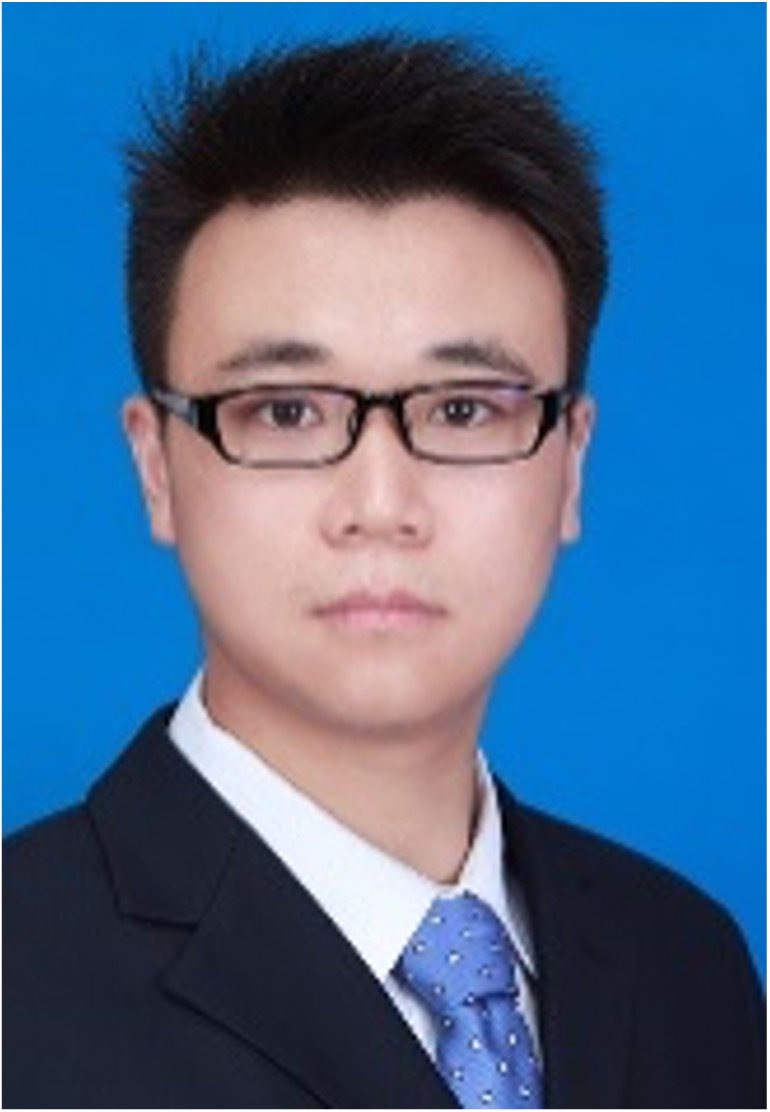}}]
{Nan Zhang} (zhang.nan152@zte.com.cn) is currently a Master Researcher and Project Manager at ZTE Corporation, Shanghai 201203, China. He received his B.S. and M.S. degrees from Tongji University in 2012 and 2015, respectively. His research interests include channel modeling, RIS, NTN, and MIMO systems. He has made numerous technical contributions to 3GPP and ETSI, and has served as a rapporteur for topics such as NR network-controlled repeaters, the 7–24 GHz channel model, and RIS.
\end{IEEEbiography}
\vskip 0pt plus -1fil
\begin{IEEEbiography}
[{\includegraphics[width=1in,height=1.25in,clip,keepaspectratio]{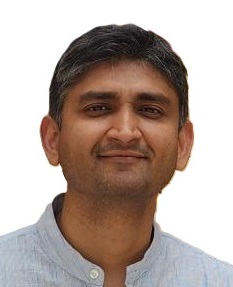}}]
{Gokul Sridharan} (gokuls@qti.qualcomm.com) Gokul Sridharan is a wireless systems engineer at Qualcomm Technologies Inc. He received his M.A.Sc. and Ph.D. degrees in Electrical Engineering from the University of Toronto, Toronto, Canada, in 2010 and 2014, respectively. His research interests include wireless communications, convex optimization and information theory and he has made contributions to the 3GPP 5G standards development. 
\end{IEEEbiography}

\vskip 0pt plus -1fil

\begin{IEEEbiography}
[{\includegraphics[width=1in,height=1.25in,clip,keepaspectratio]{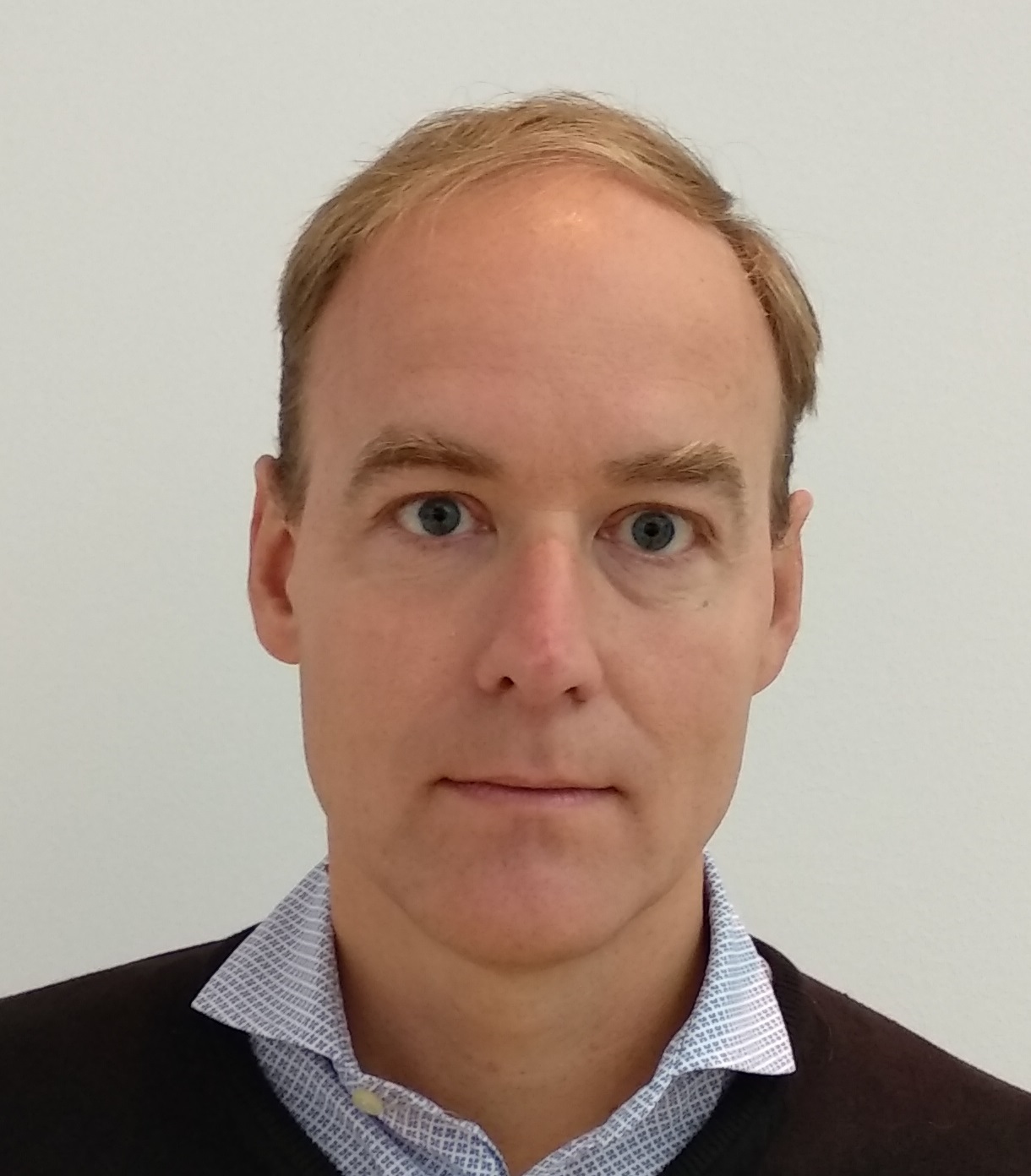}}]
{Henrik Asplund} (henrik.asplund@ericsson.com) received the MSc degree from Uppsala University, Sweden, in 1996 and joined Ericsson Research, Stockholm, Sweden, where he holds the title of Principal Researcher. He has been working in the field of antennas and propagation supporting predevelopment and standardization of all major wireless technologies from 2G to 6G. His current research interests include antenna techniques, radio channel measurements and modeling, and deployment options for 6G including higher frequencies.
\end{IEEEbiography}

\vskip 0pt plus -1fil

\begin{IEEEbiography}
[{\includegraphics[width=1in,height=1.25in,clip,keepaspectratio]{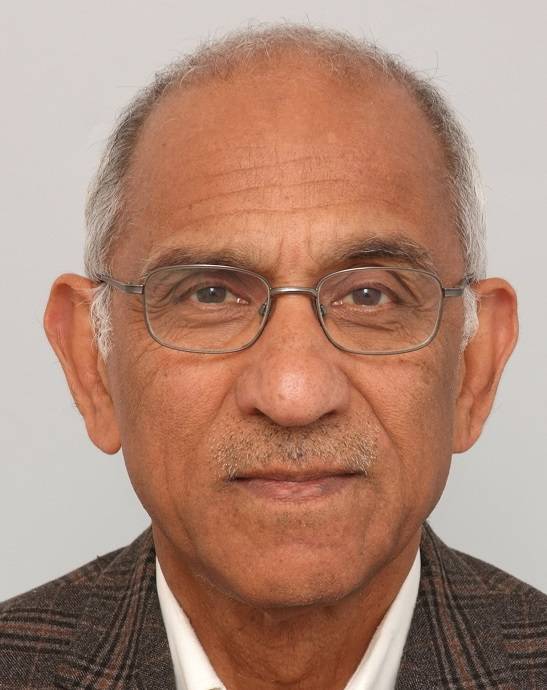}}]
{Mansoor Shafi} (mansoor.shafi@spark.co.nz) is a IEEE Life Fellow, Telecom Fellow (Wireless at Spark NZ) and an Adjunct Professor with the School of Engineering, Victoria
University of Wellington, Wellington, and the School of Engineering, University of Canterbury, Christchurch, New Zealand. He has co-authored over 200 journal and conference papers. He is a delegate to the ITU-R and 3GPP meetings about standardization of mobile radio systems.
\end{IEEEbiography}

\end{document}